\documentclass{IEEEtran}
\usepackage{cite}
\usepackage{amsmath,amssymb,amsfonts}
\usepackage{float}
\usepackage{algorithmic}
\usepackage{graphicx}
\usepackage{textcomp}
\def\BibTeX{{\rm B\kern-.05em{\sc i\kern-.025em b}\kern-.08em
    T\kern-.1667em\lower.7ex\hbox{E}\kern-.125emX}}
\begin{document}
\title{Quantum Computing Framework for Transient Scattering of Electromagnetic Waves by Dielectric Structures}
%\author{First A. Author, \IEEEmembership{Fellow, IEEE}, Second B. Author, and Third C. Author, Jr., \IEEEmembership{Member, IEEE}
\author{Min Soe, Abhay K. Ram, Efstratios Koukoutsis, George Vahala, Linda Vahala (senior member, IEEE)
, and Kyriakos Hizanidis
  \thanks{The research leading to this paper was supported by Department of Energy grants
DE-SC0021647 and DE-FG02-91ER-54109 (A.K.R.), DE-SC0021857 (M.S.), DE-SC0021651 (G.V.),
and DE-SC0021653 (L.V.).}
\thanks{M. Soe is with the Department of Mathematics, Physical Sciences and Engineering, Rogers State University,
Claremore,OK 74017 USA (e-mail: msoe.rsu@gmail.com).\\
\indent A. K. Ram is with the Plasma Science and Fusion Center, Massachusetts In-
stitute of Technology, Cambridge, MA 02139 USA (e-mail: abhay@mit.edu).\\
\indent E. Koukoutsis is with the School of Electrical \& Computer Engineering,
National Technical University of Athens, Zographou 15780 Greece (e-mail:
stkoukoutsis@mail.ntua.gr).\\
\indent G. Vahala is Emeritus with the Department of Physics, William \& Mary, Williams-
burg, VA 23185 USA (email: gvahala@gmail.com).\\
\indent L. Vahala is with the Department of Electrical \& Computer Engineer-
ing, Old Dominion University, Norfolk, VA 23529 USA (e-mail: lva-
hala@odu.edu).\\
\indent K. Hizanidis is with the School of Electrical \& Computer Engineering,
National Technical University of Athens, Zographou 15780 Greece (e-mail:
kyriakos@central.ntua.gr).}}

\maketitle

\begin{abstract}
There is an increasing expectation that quantum
computers of the near-future could simulate certain problems in
classical physics faster than traditional computers. Quantum
computers are ideally set up to solve linear systems which
are of a form similar to the Schrodinger/Dirac equation of quantum 
mechanics. This paper uses concepts from quantum information
science to craft Maxwell equations into a form suitable for
quantum computing. In the framework of linear response theory,
the propagation and scattering of electromagnetic waves in a
dielectric medium are described by Maxwell equations which
depend linearly on the electric and magnetic fields. The time evolution
 of the fields is formulated in terms of unitary operators
- a requirement which follows from the Schrodinger/Dirac formulation. This 
unitary framework is the basis for developing a qubit lattice
algorithm suitable for implementing on a quantum computer.
It consists of a series of alternating unitary streaming and entanglement operators
acting on qubit amplitudes constructed from the electric and magnetic fields.
It is not a direct discretization of Maxwell equations, but recovers the
desired equations to second order in lattice grid spacing.
The resulting algorithm is implemented
on a present-day supercomputer and is the basis of studying
scattering of electromagnetic waves by an elliptical dielectric.
As opposed to the steady state description of Mie scattering in
frequency domain, the temporal evolution provides insights
into transient scattering - a better representation of practical
and realistic scattering phenomena. The simulations, based on
the qubit lattice algorithm, reveal that a spatially localized wave
packet propagating past an elliptic dielectric, embedded in
vacuum, leads to several reflections generated by wave fields
trapped within the dielectric. The physics insight brought forth
by these simulations is not apparent from frequency domain
studies of scattering. A complimentary simulation on transient
scattering of a wave packet by an elliptical vacuum bubble
inserted in a uniform dielectric demonstrates a stark contrast with respect
to scattering off an elliptical dielectric in vacuum. Essentially,
there is only a single internal reflection in which the field
amplitudes are significantly smaller than those for side and
forward scattering. A simple model based on the Kirchhoff
tangent plane approximation helps explain
the differences between these two scattering examples.

\end{abstract}


\section{Introduction}
\label{sec:introduction}
\IEEEPARstart{T}{here} is considerable interest in devising techniques for simulating topics in classical physics
using quantum computers. The appeal of quantum computing is that it can provide exponential speed-up over classical
computers for a whole class of topics -- in particular, those that involve linear differential equations. 
Quantum computers are ideally set up to solve the linear Schr\"odinger/Dirac equation which is
a time evolution equation for a wave function describing a closed quantum process of interest. 
For electromagnetic wave propagation in a medium for which the constitutive relations are linear, the 
resulting Maxwell
equations remain linear in the fields. The constitutive relations connecting the displacement field and
the magnetic intensity to the electric field and magnetic induction, respectively, follow from linear
response theory. Thus, at first glance, we anticipate that the linear Maxwell equations are well suited for quantum
computers. The base requirement is for us to express Maxwell equations into a form that resembles the
Schr\"odinger/Dirac equation.
Following this step, we want to appropriately discretize the resulting equation in time and space for implementing on a quantum computer.
And, finally, while error correcting quantum computers with long coherence times required for reliable computations 
are not presently available, we want to implement the algorithm on contemporary classical computers and
perform some meaningful simulations related to scattering of electromagnetic waves.
These three topics form the crux of this paper.

The connection between Maxwell equations and Dirac equation for a massless particle was explored during the early days of
quantum physics \cite{lap1931,oppie1931,moses1959}. More recently, attention has turned to associating Maxwell equations with the Schr\"odinger/Dirac equation
\cite{khan, vahala-jpp, stratos-dyson}. In this paper, we will follow the procedure laid out in \cite{stratos-dyson} to express Maxwell equations
in a form resembling the Schr\"odinger/Dirac equation.
For discretizing the continuous time evolution of electromagnetic fields, we draw upon qubit lattice
algorithms which have been successful in computational analysis of a variety of quantum and classical phenomena (see \cite{vahala-jpp}
and references therein). Then the qubit lattice algorithm for the Schr\"odinger/Dirac form of Maxwell equations was
implemented in a conventional supercomputer to simulate transient scattering of a Gaussian electromagnetic wave packet by an 
elliptic cylindrical dielectric and by a similar shaped vacuum bubble.
In the former case, the dielectric is surrounded by vacuum while, in the latter case, the vacuum bubble is surrounded by a dielectric medium.

The essence of Mie scattering is to provide useful physical insight into scattering of vacuum electromagnetic waves by particles \cite{hulst, bohren, kong},
and of radio frequency waves by filamentary plasma structures \cite{ram2016, ram2013, ioann2017}. Usually, Mie scattering is studied in the frequency domain for plane electromagnetic
waves. The corresponding Maxwell equations lead to a steady state solution of the scattering process. Conceptually, we expect that a description of
scattering in the time domain will be different, especially for an electromagnetic field which is either a broadband wave packet or spatially confined pulse.
The scattering process is transient and of practical relevance as customary means of detection and
diagnostic imaging use short pulsed energy sources. As noted in \cite{kenn1958,kenn1965,benn1970,ma1992,vech1992}, 
for transient scattering it is obvious that the scattered pulse
will contain a great deal of information about the target including its shape, size, and composition.
In comparison to steady state Mie scattering, transient scattering has received limited attention, primarily theoretical modeling 
of response to an impulse function \cite{kenn1958,kenn1965,benn1970,ma1992,vech1992}. 
Our formalism including the qubit lattice algorithm is
an ideal computational test bed for transient scattering. While we graphically display the results from some of the computations,
we do not attempt comparisons with any of the previous analytical forms; this is left as an exercise for the future.

This paper is divided as follows. In section \ref{sec:2}, we introduce some basic aspects of quantum computing which are pertinent to our
desired objective. This is followed by section \ref{sec:3} on Maxwell equations and their transformation to a form resembling the Schr\"odinger/Dirac
equation. We explicitly determine the unitary operators that evolve the electromagnetic fields in time. The evolution equation is reduced to
two spatial dimensions for our computational study. 
The qubit lattice algorithm that is the basis of our computational studies is set up in section \ref{sec:4}, followed by
a pictorial representation of scattering of an electromagnetic wave packet by a 
localized dielectric and a vacuum bubble in section \ref{sec:5}. 

\section{Essential aspects of quantum computing}
\label{sec:2}

The basis of quantum computing lies in quantum physics \cite{nielson,mermin}. Some of the basic mathematical and
theoretical constructs required for studying the propagation of electromagnetic waves within the context of quantum
computing are in \cite{ram2024a}. Quantum computers are governed by the postulates of quantum mechanics which
provide a fundamental description of quantum systems.
Of the various postulates, there are two which are necessary for converting Maxwell equations
into a Schr\"odinger-like equation, and for developing a qubit lattice algorithm that discretizes the transformed equation.

The first relevant postulate: the quantum state of a system is described by a state vector $| {\psi} \rangle$\footnote{We will use the Dirac bra-ket notation
in this section, and drop it in subsequent sections.} 
which is an element in a complex Hilbert space ${\mathcal H}$. The state vector contains the complete statistical information about
the system. Just as a bit is a fundamental unit of information in a classical computer, a quantum bit, or a qubit, is a fundamental unit
of information in a quantum computer. While a bit belongs to a discrete binary system, a qubit is represented by a state vector which
is a linear superposition of a basis set spanning a two-dimensional Hilbert space ${\mathcal H}_2$,
\begin{equation}
|{\Psi} \rangle = c_0 | 0 \rangle + c_1 | 1 \rangle, \label{pos1a}
\end{equation}
where $c_0$ and $c_1$ are complex numbers.  In the so-called computational basis, the matrix representations
of the orthogonal states $| 0 \rangle $ and $| 1 \rangle $ are
\begin{equation}
| 0 \rangle = \begin{pmatrix} 1 \\ 0 \end{pmatrix}, \quad | 1 \rangle = \begin{pmatrix} 0 \\ 1 \end{pmatrix}. \label{pos1b}
\end{equation}
 It is essential that the qubit is normalized, $\langle \Psi | \Psi \rangle = 1$, so that $| c_0 |^2 + | c_1 |^2 = 1$. 
Consequently, when measuring
a qubit it collapses to a classical bit $0$ with probability $| c_0 |^2$ or a classical bit $1$ with probability $| c_1 |^2$.
The state of a qubit can be represented as a vector with its tip lying on a sphere, a Bloch sphere, centered at an origin  \cite{nielson,mermin}. 
%The
%orth pole of a Bloch sphere is the state $| 0 \rangle$ while the south pole is the state $| 1 \rangle$. Any state of a
%qubit can be written in terms of the basis set and azimuthal and polar angles, Fig \ref{fig0}.
\begin{figure}[!t]
\centerline{\includegraphics[width=0.6\columnwidth]{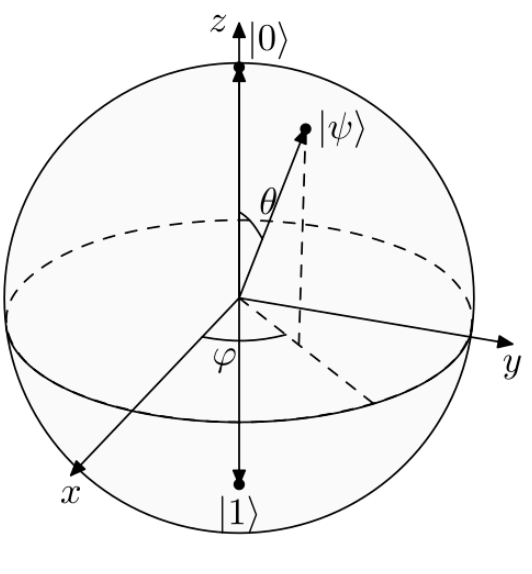}}
\caption{The Bloch sphere,  \cite{nielson,mermin}, with state $| 0 \rangle $ at the north pole and $| 1 \rangle $ at the south pole.
The state  $| \psi \rangle $ is a point on the surface of the unit sphere and can be represented in terms of this 
basis set with azimuthal and polar angles.
 }
\label{fig0}
\end{figure}

Aside from the critical advantage of linear superposition leading to simultaneous investigation of numerous possibilities
 (quantum parallelism), 
we now come to the important concept of entangled qubits.  Consider two qubits
\begin{equation}
| \Psi_1 \rangle = c_{1 0} | 0 \rangle + c_{1 1} | 1 \rangle, \quad | \Psi_2 \rangle = c_{2 0} | 0 \rangle + c_{2 1} | 1 \rangle, \label{pos1c}
\end{equation}
and their tensor product
\begin{align}
| \Psi_1 \Psi_2 \rangle  = \ & | \Psi_1 \rangle \otimes | \Psi_2 \rangle \nonumber \\
= \ &\  c_{1 0} c_{2 0} | 0 \rangle \otimes | 0 \rangle +
c_{1 0} c_{2 1} | 0 \rangle \otimes | 1 \rangle + \nonumber \\
& \ c_{1 1} c_{2 0} | 1 \rangle \otimes | 0 \rangle +
c_{1 1} c_{2 1} | 1 \rangle \otimes | 1 \rangle \nonumber \\
= \ &\  c_{1 0} c_{2 0} | 0  0 \rangle +
c_{1 0} c_{2 1} | 0  1 \rangle + \nonumber \\
& \ c_{1 1} c_{2 0} | 1  0 \rangle + c_{1 1} c_{2 1} | 1 1 \rangle , \label{pos1d}
\end{align}
where $c_{1 0}, \ c_{1 1}, \ c_{2 0}, \ c_{2 0},$ are probability amplitudes.  Since the qubits are normalized,
\begin{equation}
 |c_{1 0}|^2 + |c_{1 1}|^2 = 1,   |c_{2 0}|^2 + |c_{2 1}|^2 = 1,  \label{norm}
 \end{equation}
 and the four basis states
\begin{equation}
| 0 0 \rangle = \begin{pmatrix} 1 \\ 0 \\ 0 \\ 0 \end{pmatrix},\
| 0 1 \rangle = \begin{pmatrix} 0 \\ 1 \\ 0 \\ 0 \end{pmatrix},\
| 1 0 \rangle = \begin{pmatrix} 0 \\ 0 \\ 1 \\ 0 \end{pmatrix},\
| 1 1 \rangle = \begin{pmatrix} 0 \\ 0 \\ 0 \\ 1 \end{pmatrix}, \label{pos1e}
\end{equation}
span the four-dimensional Hilbert space ${\mathcal H}_4$.
While the tensor product $| \Psi_1 \Psi_2 \rangle \in {\mathcal H}_2 \otimes {\mathcal H}_2 \subset {\mathcal H}_4$,
there exists 2-qubit states in  ${\mathcal H}_4$ that are not tensor products of 1-qubit states, \eqref{pos1d} .
For example, consider the 2-qubit state
\begin{equation}
| \Phi \rangle = {\textsf c}_ 0 | 0 0 \rangle + {\textsf c}_1 | 1 1 \rangle, \label{pos1f}
\end{equation}
with the probability amplitudes ${\textsf c}_0$ and ${\textsf c}_1$  satisfying the normalization
$| {\textsf c}_0 |^2 + | {\textsf c}_1 |^2 = 1 $.
 It is impossible to construct the state \eqref{pos1f} for any allowed
choice of $c_{1 0}$, $c_{1 1}$, $c_{2 0}$, and $c_{2 1}$ satisfying \eqref{norm}. 
Consequently, $| \Phi \rangle \in {\mathcal H}_4 \setminus \left(  {\mathcal H}_2 \otimes {\mathcal H}_2 \right)$.  
Such states are called entangled states since in this 2-qubit state there are correlations between the individual
qubit states even when they are separated by great distances.
For ${\textsf c}_0 = {\textsf c}_1 = 1 / \sqrt{2}$, the state vector \eqref{pos1f} is an example of a maximally entangled two qubit Bell state  \cite{nielson,mermin}. 

The second relevant postulate: for a closed quantum system,
the time evolution of  $| {\psi (t)} \rangle$ is given by the Schr\"odinger/Dirac equation,
\begin{equation}
i \hbar \frac{d}{dt} | {\psi(t)} \rangle = {\mathbf H} \left( t \right) \, | {\psi(t)} \rangle, \label{sch1}
\end{equation}
where $\hbar$ is the reduced Planck constant, and
${\mathbf H}$ is a Hermitian operator, that is, ${\mathbf H}^\dagger = {\mathbf H}$ where $^\dagger$ denotes
Hermitian adjoint of the operator. Then, 
\begin{align}
\frac{d}{dt} \langle {\psi (t) | \psi (t)} \rangle & = \left( \frac{d}{dt} \langle {\psi (t)} \right) | {\psi (t)} \rangle
+ \langle {\psi (t)} \left( \frac{d}{dt} | {\psi (t)} \rangle \right) \nonumber \\
& =
\left\{ \langle{\psi (t)} | \left( \frac{1}{i \hbar} {\mathbf H} \right)^\dagger \right\} | {\psi (t)} \rangle +\nonumber \\
& \quad \quad \langle {\psi (t)} | \left\{ \left( \frac{1}{i \hbar} {\mathbf H} \right)  | {\psi (t)} \rangle \right\} \nonumber \\
& = \frac{1}{i \hbar} \langle {\psi (t)} | \left( - {\mathbf H}^\dagger + {\mathbf H} \right) | {\psi (t)} \rangle = 0. \label{sch2}
\end{align}
Thus, the Schr\"odinger/Dirac equation preserves the norm.  This is also a property of unitary operators.
If ${\mathcal U}$ is a unitary operator which maps $| {\psi (t_0)} \rangle$ at time $t_0$
to $| {\psi (t)} \rangle$ at time t,
\begin{align}
& {\mathcal U}\left( t, t_0 \right): | {\psi (t_0)} \rangle \to | {\psi (t)} \rangle \quad \Rightarrow \nonumber \\ 
& \quad | {\psi(t_0)}  \rangle 
\mapsto | {\psi(t)} \rangle = {\mathcal U} \left( t, t_0 \right) 
| {\psi(t_0)} \rangle, \label{sch3}
\end{align}
it follows that,
\begin{equation}
i \hbar \frac{d}{dt} | {\psi (t)} \rangle =i \hbar \left\{ \frac{d}{dt} {\mathcal U} \left( t, t_0 \right) \right\} \ | {\psi (t_0)} \rangle. \label{sch4}
\end{equation}
Consequently, since
\begin{equation}
i \hbar \left\{ \frac{d}{dt} {\mathcal U} \left( t, t_0 \right) \right\} \ | {\psi (t_0)} \rangle = 
 {\mathbf H} \left( t \right) \, {\mathcal U} \left( t, t_0 \right)  \ | {\psi (t_0)} \rangle. \label{sch5}
\end{equation}
for any $| { \psi (t_0)} \rangle$, we have,
\begin{equation}
i \hbar  \frac{d}{dt} {\mathcal U} \left( t, t_0 \right) = 
 {\mathbf H} \left( t \right) \, {\mathcal U} \left( t, t_0 \right). \label{sch6}
\end{equation}
If ${\mathbf H}$ is independent of time, 
\begin{equation}
{\mathcal U} \left( t, t_0 \right)  = \exp \left\{ \frac{i}{\hbar} {\mathbf H} \left( t - t_0 \right) \right\}. \label{sch7}
\end{equation}
Thus, Hermitian operators are generators of unitary operators which, in turn, evolve the quantum state in time. Accordingly,
the gates in a quantum computer are unitary operators.
\subsection{Jones vectors and the Poincar\'e sphere}
\label{sec:2a}
We can draw an analogy between the state of a qubit and polarization of electromagnetic waves in vacuum.
For a plane wave propagating along the $z$-axis in a Cartesian coordinate system, we define the $x$-direction to be along
the $s$-polarization component of the electric field ${\mathbf E}$, while the $y$-direction is along the $p$-polarization. 
The polarization states can be expressed in terms of the  Jones vectors \cite{fowles},
\begin{equation}
| 0 \rangle = \begin{pmatrix} 1 \\ 0 \end{pmatrix}, \quad | 1 \rangle  = \begin{pmatrix} 0 \\ 1 \end{pmatrix},
\label{2a1}
\end{equation}
where $ | 0 \rangle$ represents linear horizontal polarization along $x$, and $| 1 \rangle$ represents
linear vertical polarization along $y$. We can express other polarizations in terms of these Jones vectors. For example,
\begin{equation}
{\mathbf E}_{RCP} = \frac{1}{\sqrt{2}} \left( | 0 \rangle + i | 1 \rangle \right), \quad
{\mathbf E}_{LCP} = \frac{1}{\sqrt{2}} \left( | 0 \rangle - i | 1 \rangle \right), 
\end{equation}
are expressions for the right and left circularly polarized waves, respectively. 

The Poincar\'e sphere is a pictorial representation of polarized waves \cite{zang}. However, in contrast to the Bloch sphere,
the north pole on the Poincar\'e sphere represents ${\mathbf E}_{RCP}$ and the south pole represents ${\mathbf E}_{LCP}$.
The $s$- and $p$- polarizations lie in the equatorial plane with azimuthal angle of $0^\circ$ and $180^\circ$, respectively.
\section{Maxwell equations and constitutive relations}
\label{sec:3}
%section 3

For electromagnetic wave propagation in a medium, the Maxwell equations are,
\begin{align}
\nabla \pmb{\cdot}  {\mathbf D} \left( {\mathbf r}, t \right) \ &= \ 0, \label{3.1} \\
% \nabla \pmb{\cdot} \left\{ \epsilon \left( {\mathbf r} \right) \; {\mathbf E} \left( {\mathbf r}, t \right) \right\} \ & = \ 0, \\
\nabla \pmb{\cdot} {\mathbf B} \left( {\mathbf r}, t \right) \ &= \ 0,  \label{3.2} \\
\nabla \times {\mathbf E} \left( {\mathbf r}, t \right) \ &= \ -\, \frac{\partial {\mathbf B} \left( {\mathbf r}, t \right)}{\partial t}, \label{3.3}\\
 \nabla \times {\mathbf H} \left( {\mathbf r}, t \right)\ &= \ \frac{\partial {\mathbf D} \left( {\mathbf r}, t \right)}{\partial t}. \label{3.4}
\end{align}
where we have assumed that there are no external charges or currents.
In our notation, ${\mathbf E}$ is the electric field, ${\mathbf D}$ is the displacement electric field, 
${\mathbf B}$ is the magnetic induction or magentic field, ${\mathbf H}$ is the magnetic intensity. 
The Gauss law \eqref{3.1} and the Gauss magnetization law \eqref{3.2} can be considered as initial conditions. Once these two laws are satisfied at initial
time $t = 0$, they are satisfied for all times. This can be proved by taking the divergence of%of the Ampere-Maxwell 
\eqref{3.3} and 
 of  \eqref{3.4}. As a result, we usually solve the time evolution equations \eqref{3.3} and \eqref{3.4} 
simultaneously. Since these are two vector
equations for four field vectors ${\mathbf E}$, $\mathbf D$, $\mathbf B$, and $\mathbf H$ the system of Faraday-Ampere-Maxwell equations is
indeterminate. Here we introduce constitutive relations between $\mathbf D$ and $\mathbf E$ and between $\mathbf H$ and $\mathbf B$. The
constitutive relations are determined by physics external to Maxwell equations, and describe the response of a medium to the 
electromagnetic fields $\mathbf E$ and $\mathbf B$. We will assume that the response of the medium is linear in $\mathbf E$ and 
$\mathbf B$ and local in space, and the medium is a non-dispersive, non-dissipative dielectric \cite{zang}. Furthermore, the response of
the medium does not vary with time.
Then,
\begin{equation}
{\mathbf D} \left( {\mathbf r}, t \right)= {\pmb \epsilon} \left( {\mathbf r} \right) \pmb{\cdot} {\mathbf E} \left( {\mathbf r}, t \right), \quad
{\mathbf H} \left( {\mathbf r}, t \right)= \frac{1}{\mu_0} \, {\mathbf B} \left( {\mathbf r}, t \right), \label{3.5}
\end{equation}
where ${\pmb \epsilon} \left( {\mathbf r} \right)$ is the spatially dependent permittivity of the medium which, in general,
is a three-dimensional rank-2 tensor, and $\mu_0$ is the permeability of vacuum. 
%We will assume that there are no absorption losses in the dielectric medium.
Consequently, energy in the electromagnetic wave is conserved and the permittivity tensor is therefore
Hermitian. Since a Hermitian tensor can be diagonalized,
we can choose a coordinate system defined by the principal directions corresponding to the three eigenvalues. Without loss of generality, we
choose the coordinate system to be Cartesian and express the permittivity tensor as,
\begin{equation}
\pmb{\epsilon} = \begin{pmatrix} 
\epsilon_x \left( {\mathbf r} \right)& 0 & 0 \\
0 & \epsilon_y \left( {\mathbf r} \right) & 0  \\
0 & 0 & \epsilon_z \left( {\mathbf r} \right)
\end{pmatrix}. \label{3.6}
\end{equation}
\subsection{Schr\"odinger/Dirac  form of Maxwell equations}
%section 3a

In framing Faraday-Ampere-Maxwell equations into a Schr\"odinger equation, we will follow the formalism outlined in
\cite{stratos-dyson}. We define two column vectors in the Cartesian $\left( x, y, z \right)$ coordinate system,
\begin{equation}
{\mathbf d} = \begin{pmatrix} D_x \\ D_y \\ D_z \\ B_x \\ B_y \\ B_z \end{pmatrix}, \quad
{\mathbf u} = \begin{pmatrix} E_x \\ E_y \\ E_z \\ H_x \\ H_y \\ H_z \end{pmatrix}. \label{3a1}
\end{equation}
The column vectors are related through the constitutive relations,
\begin{equation}
{\mathbf d} = \widehat{\mathbf W} \pmb{\cdot} {\mathbf u}, \quad {\rm where} \ \ \widehat{\mathbf W} = \begin{pmatrix} 
\pmb{\epsilon} \left( {\mathbf r} \right) & 0_{3 \times 3} \\
0_{3 \times 3} & \mu_0 {\mathbf I}_{3 \times 3}
\end{pmatrix}, \label{3a2}
\end{equation}
and $0_{3 \times 3}$ and ${\mathbf I}_{3 \times 3}$ are rank-2, three-dimensional, null and identity tensors, respectively.
Then,  \eqref{3.3} and \eqref{3.4} take on the form,
\begin{equation}
i \frac{\partial}{\partial t} {\mathbf d}  = \widehat{\mathbf M} \pmb{\cdot} {\mathbf u}, \label{3a3}
\end{equation}
with,
\begin{equation}
\widehat{{\mathbf M}} = \begin{pmatrix} 
0_{3 \times 3} & i \widehat{\pmb{\Delta}} \\[3pt]
- i \widehat{\pmb{\Delta}} & 0_{3 \times 3}  
\end{pmatrix} \ \
{\rm and} \ \ 
\widehat{\pmb{\Delta}} = \begin{pmatrix}
0 & -\frac{\partial}{\partial z} & \frac{\partial}{\partial y} \\[6pt]
\frac{\partial}{\partial z} & 0 & -\frac{\partial}{\partial x} \\[6pt]
-\frac{\partial}{\partial y} & \frac{\partial}{\partial x} & 0
\end{pmatrix}. \label{3a4}
\end{equation}
The operator $i \widehat{\pmb \Delta}$ is Hermitian with the appropriate boundary conditions (usually, Dirichlet or Neumann) 
. \footnote{ $i \frac{\partial}{\partial x}$, $i \frac{\partial}{\partial y}$, and $i \frac{\partial}{\partial z}$ are
self-adjoint differential operators.} As a result, $\widehat{\mathbf M}$ is a Hermitian operator.
From \eqref{3a2} and \eqref{3a3},
\begin{equation}
i \frac{\partial}{\partial t} {\mathbf u}  = \widehat{\mathbf W}^{-1} \pmb{\cdot} \widehat {\mathbf M} \pmb{\cdot} {\mathbf u}. \label{3a5}
\end{equation} 
In general, the operator $\widehat{\mathbf W}^{-1} \pmb{\cdot}  \widehat{\mathbf M}$ is not Hermitian; the spatial
derivatives do not commute through the spatially dependent permittivity tensor. As a consequence, the time evolution of
${\mathbf u}$ in \eqref{3a5} cannot be expressed in terms of a unitary operator. It is only in the special case of a homogeneous
dielectric medium that $\widehat{\mathbf W}^{-1} \pmb{\cdot}  \widehat{\mathbf M}$ is Hermitian.
\subsection{Formulating unitary evolution of Maxwell equations}
%section 3b

The conversion of the generator $\widehat{\mathbf W}^{-1} \pmb{\cdot}  \widehat{\mathbf M}$ in \eqref{3a5} to a Hermitian operator
is achieved by transforming to a new set of electromagnetic field variables. The existence of such a transformation, referred to as a Dyson map,
has been discussed in \cite{stratos-dyson}. The transformed fields are given by,
\begin{align}
{\mathbf Q} & =\widehat {\mathbf W}^{1/2} \pmb{\cdot} {\mathbf u} \nonumber  
 = 
%\begin{pmatrix}
%\sqrt{\epsilon_x} & 0 & 0 & 0 & 0 & 0 \\
%0 & \sqrt{\epsilon_y} & 0 & 0 & 0 & 0  \\
%0 & 0 & \sqrt{\epsilon_z} & 0 & 0 & 0  \\
%0 & 0 & 0 & \sqrt{\mu_0} & 0 & 0 \\
%0 & 0 & 0 & 0 & \sqrt{\mu_0} & 0 \\
%0 & 0 & 0 & 0 & 0 & \sqrt{\mu_0}
%\end{pmatrix} \pmb{\cdot} {\mathbf u} \nonumber \\
%& = 
\begin{pmatrix}
\sqrt{\epsilon_x}\, E_x \\
\sqrt{\epsilon_y} \, E_y \\
\sqrt{\epsilon_z} \, E_z \\
\sqrt{\mu_0} \, H_x \\
\sqrt{\mu_0} \, H_y \\
\sqrt{\mu_0} \, H_z \\
\end{pmatrix}. 
\label{3b1}
\end{align}
From \eqref{3a5}, we obtain the evolution equation for $\mathbf Q$,
\begin{equation}
i \frac{\partial}{\partial t} {\mathbf Q}  = \widehat{\mathbf W}^{-1/2} \pmb{\cdot} \widehat {\mathbf M} \pmb{\cdot} 
\widehat{\mathbf W}^{-1/2}\pmb{\cdot} {\mathbf Q}. \label{3b2}
\end{equation}
At this stage, $\widehat{\mathbf W}^{-1/2} \pmb{\cdot} \widehat {\mathbf M} \pmb{\cdot} 
\widehat{\mathbf W}^{-1/2}$ is Hermitian, and the evolution of ${\mathbf Q}$ in time is prescribed by a unitary operator.

If we multiply ${\mathbf Q}$ with $c \sqrt{\mu_0}$, where $c$ is the speed of light,
\begin{equation}
{\mathbf Q} \rightarrow 
\begin{pmatrix}
n_x\, E_x \\
n_y \, E_y \\
n_z \, E_z \\
c \, B_x \\
c \, B_y \\
c \, B_z \\
\end{pmatrix}
\ \equiv \ 
\begin{pmatrix}
q_0 \\
q_1 \\
q_2 \\
q_3 \\
q_4 \\
q_5 
\end{pmatrix}. \label{3b3}
\end{equation} 
with $n_x = c \sqrt{ \epsilon_x \mu_0}$, $n_y = c \sqrt{ \epsilon_y \mu_0}$, and $n_z = c \sqrt{ \epsilon_z \mu_0}$, being the
indicies  of refraction along the three principal directions. The indicies are, in general, functions of space.
We have denoted the elements of $\mathbf Q$ by $q$'s for subsequent
convenience in setting up the numerical algorithm; we will refer to $q$'s as qubit amplitude. 
If we multiply \eqref{3b2} by $1/c$ and replace $ct \rightarrow t$,\footnote{Thus, $t$ has the dimension of space.} then the equation
retains its mathematical form with ${\mathbf Q}$ as in \eqref{3b3}, and the operator on the right-hand side given by,
\newpage
\begin{multline}
\widehat{\mathbf W}^{-1/2} \pmb{\cdot} \widehat {\mathbf M} \pmb{\cdot} 
\widehat{\mathbf W}^{-1/2} \ = \ \\
\begin{pmatrix}
0 & 0 & 0 & 0 & - \frac{i}{n_x} \partial_z & \frac{i}{n_x} \partial_y \\[5pt]
0 & 0 & 0 & \frac{i}{n_y} \partial_z & 0 & - \frac{i}{n_y} \partial_x \\[5pt]
0 & 0 & 0 & - \frac{i}{n_z} \partial_y & \frac{i}{n_z} \partial_x & 0 \\[5pt]
0 & i \partial _z \frac{1}{n_y} & -i \partial_y \frac{1}{n_z} & 0 & 0 & 0 \\[5pt]
- i \partial_z \frac{1}{n_x} & 0 & i \partial_x \frac{1}{n_z} & 0 & 0 & 0 \\[5pt]
i \partial_y \frac{1}{n_x} & -i \partial_x \frac{1}{n_y} & 0 & 0 & 0 & 0
\end{pmatrix},\\
\label{3b4}
\end{multline}
where $\partial_x = \partial / \partial x$ and similarly for $\partial_y$ and $\partial_z$. The
spatial derivatives operate on any expressions to the right of them. The evolution equation \eqref{3b2} along
with \eqref{3b3} and \eqref{3b4} form the basis for studying the propagation and scattering of electromagnetic 
waves in three-dimensional, spatially varying dielectric media.

\subsection{Conservation of instantaneous total electromagnetic energy}
Because of the unitary evolution in the Dyson variables, the norm of the Dyson variables must be a constant 
of the evolution.  This constant is immediately seen to be the total electromagnetic energy $\mathcal{E}(t) = 
|| Q ||^2 = const,$ with
\begin{align} 
\mathcal{E}(t) =  \int_0^L \int_0^L dx dy \left[ n_x^2 E_x^2 + n_y^2 E_y^2 + n_z^2 E_z^2 )+  c^2 
\mathbf{B}^2 \right]  \label{eqA13} \\
\noindent = const \qquad \qquad . \qquad \qquad   
\end{align}  
\subsection{Evolution equations for a two-dimensional inhomogeneous medium}
If the spatial variation of the permittivity is in the $x$ and $y$ directions only, the evolution equations for each
component of $\mathbf Q$ are \cite{vah2021a,vah2021b},
\begin{align}
\partial_t q_0 \ &= \ \frac{1}{n_x} \, \partial_y q_5 \label{2d1} \\
\partial_t q_1 \ &= \ - \frac{1}{n_y} \, \partial_x q_5 \label{2d2} \\
\partial_t q_2 \ &= \ - \frac{1}{n_z} \, \partial_y q_3 + \frac{1}{n_z} \, \partial_x q_4 \label{2d3} \\
\partial_t q_3 \ &=
\ - \frac{1}{n_z} \, \partial_y  q_2 + \frac{1}{n_z^2} q_2 \, \partial_y n_z \label{2d4} \\
\partial_t q_4 \ 
&= \ \frac{1}{n_z} \partial_x \, q_2 - \frac{1}{n_z^2} q_2 \, \partial_x  n_z \label{2d5} \\
\partial_t q_5 \ 
&= \ \frac{1}{n_x} \, \partial_y q_0 - \frac{1}{n_y} \, \partial_x q_1
- \frac{1}{n_x^2} q_0 \, \partial_y n_x  + \frac{1}{n_y^2} q_1 \, \partial_x n_y. \label{2d6}
\end{align}
From \eqref{2d1}-\eqref{2d3}, it is noticeable that the time evolution equations of $q_0$, $q_1$, and $q_2$ are not explicitly
dependent of the spatial inhomogeneity gradients in the medium.
\section{Qubit Lattice Algorithm}
\label{sec:4}
We now develop a qubit lattice algorithm (QLA) for the numerical solution of the 
qubit amplitude equations \eqref{2d1}-\eqref{2d6}.  QLA consists of an interleaved
 sequence of unitary collide-stream operators
acting on qubit amplitudes located on a spatial grid of a square lattice.  
The unitary collision operator entangles qubits at a given grid point \footnote{In some 
instances, we refer to the collision operator as an entanglement operator.}
, while the unitary streaming operator
moves this entanglement throughout the lattice.  The aim is to recover a second order accurate spatial scheme
in the time advancement of the qubit amplitudes.  
%%%%%%%%%%%%%%%%%%%%%%%%%%%%%%%  addition April 18, 2026
QLA was initialized designed for solving the linear Schrodinger equation \cite{bogh1998,yepez2001} and then Burgers
equation \cite{yepez2002}.  It was then further benchmarked by simulating the exactly soluble
one-dimensional solition-soliton collisons for the nonlinear Schrodinger and Manakov 
equations \cite{vah2003b,vah2004,vah2005}.  This then permitted simulations of two- and three-dimensional quantum 
turbulence in Bose-Einstein condensates \cite{yep2009a,yep2009b,boz2011,vah2011a,vah2012,vah2012a,lvah2019a,
lvah2019b,vah2020a,vah2020b}.  Recent efforts have concentrated on wave propagation in Maxwell equations
\cite{vahala-jpp,stratos-dyson,vah2021a,vah2021b,gvaha2022, gvaha2023a, gvaha2023b, stratos2023,stratos2024,
stratos2025,soe2026a,soe2026b}.

Since we are working on a Cartesian grid, we have operator
splitting so that the $x$ and $y$ variations can be treated independently.  Let $\delta$ be the spatial lattice grid
spacing.  For some problems, it is not straightforward to fully recover a total unitary QLA representation for the
 qubit amplitude equations \eqref{2d1}-\eqref{2d6}.  In these cases, one must 
also introduce appropriate potential operators that are non-unitary.
Using operator splitting, we first consider x-variations - and just x-variations in the qubit amplitudes.  In this case, \eqref{2d2}, \eqref{2d6},
\eqref{2d3}, and \eqref{2d5}  are, respectively,
\begin{align}
\partial_t q_1 &= -\frac{1}{n_y} \, \partial_x q_5,  &\quad \partial_t q_5 &= -\frac{1}{n_y} \, \partial_x q_1, \label{2e1} \\
\partial_t q_2 &= \frac{1}{n_z} \, \partial_x q_4,   &\quad \partial_t q_4 &= \frac{1}{n_z} \, \partial_x q_2. \label{2e2}
\end{align}

From \eqref{2e1}, we note that there is a direct link between the time evolutions of $q_1$ and $q_5$ -- they are Similarly, $q_2$ and $q_4$ in \eqref{2e2} are entangled at a fixed grid point. 
We thus introduce the unitary local collision operator,
\begin{equation}
\widehat {\mathbf C}_{\rm x} = \begin{pmatrix} 
1 & 0 & 0 & 0 & 0 & 0\\
0 & \cos \theta_1& 0 & 0 & 0 & -\sin \theta_1 \\
0 & 0 & \cos \theta_2 & 0 & -\sin \theta_2 & 0\\
0 & 0 & 0 & 1 & 0 & 0\\
0 & 0 & \sin \theta_2 & 0 & \cos \theta_2 & 0\\
0 & \sin \theta_1 & 0 & 0 & 0 & \cos \theta_1 
\end{pmatrix}, \label{2e3}
\end{equation}
with,
\begin{equation}
\theta_1 =\frac{1}{4 n_y} \delta, \quad \theta_2 = \frac{1}{4 n_z} \delta, \label{2e4}
\end{equation}
where $\delta$ is the spacing between neighboring grid points. The angles $\theta_1$ and $\theta_2$ lead to 
the coefficients multiplying the spatial derivatives in \eqref{2e1} and \eqref{2e2}. 
%The operator $\widehat{\mathbf C}_{\rm x}$ is unitary.

The unitary streaming operators for the system \eqref{2e1}-\eqref{2e2} will either stream the 
amplitude pairs \{$q_, q_4$\} or \{$q_2, q_5$\}: \,  i.e., we consider the unitary streaming operators 
$\widehat{\mathbf S}_{14}^{\pm {\rm x}}$ and $\widehat{\mathbf S}_{25}^{\pm {\rm x}}$ defined by
\begin{align}
\widehat{\mathbf S}_{14}^{\pm {\rm x}} \pmb{\cdot} 
\begin{pmatrix}
q_0 \left( x, y \right) \\
q_1 \left( x, y \right) \\
q_2 \left( x, y \right) \\
q_3 \left( x, y \right) \\
q_4 \left( x, y \right) \\
q_5 \left( x, y \right) 
\end{pmatrix} \ &= \ 
\begin{pmatrix}
q_0 \left( x, y \right) \\
q_1 \left( x \pm \delta, y \right) \\
q_2 \left( x, y \right) \\
q_3 \left( x, y \right) \\
q_4 \left( x \pm \delta, y \right) \\
q_5 \left( x, y \right) 
\end{pmatrix}, \nonumber \\
\widehat{\mathbf S}_{25}^{\pm {\rm x}} \pmb{\cdot} 
\begin{pmatrix}
q_0 \left( x, y \right) \\
q_1 \left( x, y \right) \\
q_2 \left( x, y \right) \\
q_3 \left( x, y \right) \\
q_4 \left( x, y \right) \\
q_5 \left( x, y \right) 
\end{pmatrix} \ &= \ 
\begin{pmatrix}
q_0 \left( x, y \right) \\[4pt]
q_1 \left( x, y \right) \\
q_2 \left( x  \pm \delta, y \right) \\
q_3 \left( x, y \right) \\
q_4 \left( x, y \right) \\
q_5 \left( x \pm \delta, y \right) 
\end{pmatrix}. \label{2e5}
\end{align}

We now define the evolution operator $\widehat{\mathbf U}_{\rm x}(\delta)$  along the $x$-direction as the
 following interleaved sequence of non-commuting unitary collision and streaming operators 
 \begin{align}
\widehat{\mathbf U}_{\rm x} (\delta)=\ &
\widehat{\mathbf S}_{25}^{+ {\rm x}} \, \widehat{\mathbf C}_{\rm x}^\dagger \,
\widehat{\mathbf S}_{25}^{- {\rm x}} \, \widehat{\mathbf C}_{\rm x} \pmb{\cdot}
\widehat{\mathbf S}_{14}^{- {\rm x}} \, \widehat{\mathbf C}_{\rm x}^\dagger  \,
\widehat{\mathbf S}_{14}^{+ {\rm x}} \, \widehat{\mathbf C}_{\rm x} \pmb{\cdot} \nonumber \\
& \widehat{\mathbf S}_{25}^{- {\rm x}} \, \widehat{\mathbf C}_{\rm x} \,
\widehat{\mathbf S}_{25}^{+ {\rm x}} \,\widehat{\mathbf C}_{\rm x}^\dagger \pmb{\cdot}
\widehat{\mathbf S}_{14}^{+ {\rm x}} \, \widehat{\mathbf C}_{\rm x} \,
\widehat{\mathbf S}_{14}^{- {\rm x}} \, \widehat{\mathbf C}_{\rm x}^\dagger . \label{2e6}
\end{align}
Notice that $\widehat{\mathbf U}_{\rm x}(\delta)$ can be grouped into a product of 4 composite sets.  
If the collision and streaming operators in each set commuted, then that set would be nothing but the
identity operator (since the collision operator is unitary:  $\widehat{\mathbf C}_{\rm x}^\dagger \cdot 
\widehat{\mathbf C}_{\rm x} = I $).
Hence the evolution operator $\widehat{\mathbf U}_{\rm x}(\delta)$ is a perturbation away from the identity
operator.  Let ${\mathbf Q}_{x} \left( t \right)$ be the qubit amplitude vector at time t at grid point $x$ for fixed $y$.
Using a symbolic manipulator like $\it{Mathematica}$ we find perturbatively
\begin{equation}
 \widehat{\mathbf U}_{\rm x} (\delta)\pmb{\cdot} {\mathbf Q}_x \left( t \right) = {\mathbf Q}_x \left( t \right) + 
 \delta^2 . {\mathbf P_x} + O( \delta^4 )
\end{equation}
for some vector ${\mathbf P_x}$ that involves the $x$-derivative  of the qubit amplitudes  in
\eqref{2e1}-\eqref{2e2}.  Thus the time evolution of the qubit amplitude vector in time step $\Delta t$ is
\begin{align}
{\mathbf Q}_{x+\delta} \left( t + \Delta t \right) = \widehat{\mathbf U}_{\rm x}(\delta) \pmb{\cdot} {\mathbf Q}_x \left( t \right) \qquad  \qquad \\
  = {\mathbf Q}_x \left( t \right) +  \delta^2 . {\mathbf P} + O( \delta^4 )  .
\label{2e7}
\end{align}
Hence our QLA scheme will perturbatively recover \eqref{2e1}-\eqref{2e2} to second-order in the lattice spatial
grid spacing $\delta$ under diffusion time ordering $\Delta t = \delta^2$.

Following the same procedure for evolution along the $y$-axis, we note the coupling between qubit amplitude
pairs $\{q_0,q_5\}$ and $\{q_2,q_3\}$.  Hence we introduce  the unitary entanglement operator,
\begin{equation}
\widehat {\mathbf C}_{\rm y} = \begin{pmatrix} 
\cos \theta_0 & 0 & 0 & 0 & 0 & \sin \theta_0 \\
0 & 1 & 0 & 0 & 0 & 0 \\
0 & 0 & \cos \theta_2 & \sin \theta_2 & 0 & 0\\
0 & 0 & -\sin \theta_2 & \cos \theta_2 & 0 & 0\\
0 & 0 & 0 & 0 & 1 & 0\\
-\sin \theta_0 & 0 & 0 & 0 & 0 & \cos \theta_0 
\end{pmatrix}, \label{2e8}
\end{equation}
where,
\begin{equation}
\theta_0 \ = \ \frac{1}{4 n_x} \delta,
\end{equation}
and the operator entangles the qubit pairs $\left( q_0, q_5 \right)$ and $\left( q_2, q_3 \right)$. The corresponding
streaming operators are $\widehat{\mathbf S}_{0 3}^{\pm {\rm y}}$ and $\widehat{\mathbf S}_{2 5}^{\pm {\rm y}}$, 
and the unitary evolution operator is,
\begin{align}
\widehat{\mathbf U}_{\rm y}(\delta) = \ &
\widehat{\mathbf S}_{25}^{+ {\rm y}} \, \widehat{\mathbf C}_{\rm y}^\dagger \,
\widehat{\mathbf S}_{25}^{- {\rm y}} \, \widehat{\mathbf C}_{\rm y} \pmb{\cdot}
\widehat{\mathbf S}_{03}^{- {\rm y}} \, \widehat{\mathbf C}_{\rm y}^\dagger  \,
\widehat{\mathbf S}_{03}^{+ {\rm y}} \, \widehat{\mathbf C}_{\rm y} \pmb{\cdot} \nonumber \\
& \widehat{\mathbf S}_{25}^{- {\rm y}} \, \widehat{\mathbf C}_{\rm y} \,
\widehat{\mathbf S}_{25}^{+ {\rm y}} \, \widehat{\mathbf C}_{\rm y}^\dagger \pmb{\cdot}
\widehat{\mathbf S}_{03}^{+ {\rm y}} \, \widehat{\mathbf C}_{\rm y} \,
\widehat{\mathbf S}_{03}^{- {\rm y}} \, \widehat{\mathbf C}_{\rm y}^\dagger . \label{2e9}
\end{align}
As for evolution along the $x$-axis, $\widehat{\mathbf U}_{\rm y}(\delta)$ is a perturbation from the identity operator.

For the simple special case of spatially homogeneous 2D dielectric medium,
the fully unitary QLA evolution of the state qubit vector is given by,
\begin{equation}
{\mathbf Q} \left( t + \Delta t \right) = \widehat{\mathbf U}_{\rm y}(\delta) \pmb{\cdot} \widehat{\mathbf U}_{\rm x} (\delta)
\pmb{\cdot} {\mathbf Q} \left( t \right). \label{2e10}
\end{equation}
and recovers the continuum Maxwell equations \eqref{2e1} and \eqref{2e2} to $O(\delta^2)$.
It is essential to note that \eqref{2e10} is not a discretization of
\eqref{2e1} and \eqref{2e2} in the usual sense of expressing derivatives through some form of finite differences. 
The QLA is just a set of 1-qubit, 2-qubit unitary gates and is immediately encodable on a near future quantum computer.

\subsection{Maxwell equations in inhomogeneous dielectric media}
We must now extend the simple unitary QLA \eqref{2e10} to incorporate spatial derivatives on the refractive indices and 
solve \eqref{2d1}-\eqref{2d6} for 2D dielectric media.  (Because of operator splitting it is trivial to include 
inhomogneities in the $z$-direction)

In order to include the effects of the spatial derivative of the indices of refraction, we have to introduce two
potential operators corresponding to the $x$ and $y$ derivatives, respectively,
\begin{equation}
\widehat{\mathbf V}_{\rm x}(\delta) = \begin{pmatrix} 
1 & 0 & 0 & 0 & 0 & 0 \\
0 & 1 & 0 & 0 & 0 & 0 \\
0 & 0 & 1 & 0 & 0 & 0\\
0 & 0 & 0 & 1 & 0 & 0\\
0 & 0 & -\sin \beta_2  & 0 & \cos \beta_2 & 0\\
0 & -\sin \beta_0 & 0 & 0 & 0 & \cos \beta_0 
\end{pmatrix}, \label{2e11}
\end{equation}
\begin{equation}
\widehat{\mathbf V}_{\rm y}(\delta) = \begin{pmatrix} 
1 & 0 & 0 & 0 & 0 & 0 \\
0 & 1 & 0 & 0 & 0 & 0 \\
0 & 0 & 1 & 0 & 0 & 0\\
0 & 0 & \cos \beta_3 & \sin \beta_3 & 0 & 0\\
0 & 0 & 0  & 0 & 1 & 0\\
-\sin \beta_1 & 0 & 0 & 0 & 0 & \cos \beta_1 
\end{pmatrix}, \label{2e12}
\end{equation}
with,
%\begin{align}
%\beta_0 & = \frac{1}{n_y^2} \frac{\partial n_y}{\partial x} \delta^2,  \quad  
%\beta_1  = \frac{1}{n_x^2} \frac{\partial n_x}{\partial y} \delta^2, \nonumber \\
%\beta_2 & = \frac{1}{n_z^2} \frac{\partial n_z}{\partial x} \delta^2, \quad  
%\beta_3  = \frac{1}{n_z^2} \frac{\partial n_z}{\partial y} \delta^2.  \label{2e13}
%\end{align} 
\begin{equation}
  \begin{split}
\beta_0  = \frac{1}{n_y^2} \frac{\partial n_y}{\partial x} \delta^2,  \quad  
\beta_1  = \frac{1}{n_x^2} \frac{\partial n_x}{\partial y} \delta^2,  \\
\beta_2  = \frac{1}{n_z^2} \frac{\partial n_z}{\partial x} \delta^2, \quad  
\beta_3  = \frac{1}{n_z^2} \frac{\partial n_z}{\partial y} \delta^2.  \label{2e13}
  \end{split}
\end{equation}
These two potential operators  $\widehat{\mathbf V}_{\rm x}(\delta) $ and 
$\widehat{\mathbf V}_{\rm y}(\delta)$ are not unitary
The full QLA for an inhomogeneous dielectric medium is,
\begin{equation}
{\mathbf Q} \left( t + \Delta t \right) = \widehat{\mathbf V}_{\rm y}(\delta)
 \pmb{\cdot} \widehat{\mathbf V}_{\rm x}(\delta) \pmb{\cdot}
\widehat{\mathbf U}_{\rm y}(\delta) \pmb{\cdot} \widehat{\mathbf U}_{\rm x}(\delta)
\pmb{\cdot} {\mathbf Q} \left( t \right). \label{2e14}
\end{equation}
The discretized QLA system \eqref{2e14} recovers the Maxwell equations for wave propagation in a two-dimensional inhomogeneous dielectric
medium to $O(\delta^2)$.

The potential operators in \eqref{2e11} and \eqref{2e12} are not unitary. While this does not limit the development and implementation
of our algorithm on classical computers, the formalism needs to be reframed for quantum computers. This is the subject of our next
section.
\subsection{Contending with non-unitary operators for quantum computers -- linear combination of unitaries (LCU) 
\cite{childs2012} }
It is well known that any square matrix can always be written as a sum of unitary matrices.  In particular, we can readily find 4 unitary matrices whose sum yield the potential matrices \eqref{2e11} and \eqref{2e12}:

\begin{equation}
 \widehat{\mathbf V}_{\rm x}(\delta) = \frac{1}{2} \, \sum_{i=1}^{4} \, \widehat{\mathbf L}_{{\rm x} i}, \quad
 \widehat{\mathbf V}_{\rm y}(\delta) = \frac{1}{2} \, \sum_{i=1}^{4} \, \widehat{\mathbf L}_{{\rm y} i}, 
\label{2f1}
\end{equation}
where 
\begin{equation}
\widehat{\mathbf L}_{{\rm x} 1} = {\mathbf I}_{6\times6},
\end{equation}

\begin{equation}
\widehat{\mathbf L}_{{\rm x}2} = \begin{pmatrix} 
-1 & 0 & 0 & 0 & 0 & 0\\
0 & 1 & 0 & 0 & 0 & 0\\
0 & 0 & 1 & 0 & 0 & 0\\
0 & 0 & 0 & -1 & 0 & 0\\
0 & 0 & 0 & 0 & -1 & 0\\
0 & 0 & 0 & 0 & 0 & -1 
\end{pmatrix}, \label{2f2}
\end{equation}

\begin{equation}
\widehat{\mathbf L}_{{\rm x}3} = \begin{pmatrix} 
1 & 0 & 0 & 0 & 0 & 0\\
0 & \cos \beta_0  & 0 & 0 & 0 & - \sin \beta_0\\
0 & 0 & \cos \beta_2 & 0 & \sin \beta_2 & 0\\
0 & 0 & 0 & 1 & 0 & 0\\
0 & 0 & -\sin \beta_2 & 0 & \cos \beta_2 & 0\\
0 & \sin \beta_0 & 0 & 0 & 0 & \cos \beta_0 
\end{pmatrix}, \label{2f3}
\end{equation}

\begin{equation}
\widehat {\mathbf L}_{{\rm x}4} = \begin{pmatrix}  
1 & 0 & 0 & 0 & 0 & 0\\ 
 0 & -\cos \beta_0 & 0 & 0 & 0 & \sin \beta_0\\
0 & 0 & -\cos \beta_2 & 0 & -\sin \beta_2 & 0\\
0 & 0 & 0 & 1 & 0 & 0\\
0 & 0 & -\sin \beta_2 & 0 & \cos \beta_2 & 0\\
0 & \sin \beta_0 & 0 & 0 & 0 & \cos \beta_0 
\end{pmatrix}, \label{2f4}
\end{equation}

\begin{equation}
\widehat{\mathbf L}_{{\rm y} 1} = {\mathbf I}_{6\times6},
\end{equation}

\begin{equation}
\widehat{\mathbf L}_{{\rm y}2} = \begin{pmatrix} 
1 & 0 & 0 & 0 & 0 & 0\\
0 & -1 & 0 & 0 & 0 & 0\\
0 & 0 & 1 & 0 & 0 & 0\\
0 & 0 & 0 & -1 & 0 & 0\\
0 & 0 & 0 & 0 & -1 & 0\\
0 & 0 & 0 & 0 & 0 & -1 
\end{pmatrix}, \label{2f5}
\end{equation}

\begin{equation}
\widehat{\mathbf L}_{{\rm y}3} = \begin{pmatrix} 
\cos \beta_1 & 0 & 0 & 0 & 0 & \sin \beta_1 \\
0 & 1  & 0 & 0 & 0 & 0 \\
0 & 0 & -\sin \beta_3 & \cos \beta_3 & 0 & 0 \\
0 & 0 & \cos \beta_3 & \sin \beta_3 & 0 & 0 \\ 
0 & 0 & 0 & 0 & 1 & 0 \\
-\sin \beta_1 & 0 & 0 & 0 & 0 & \cos \beta_1 
\end{pmatrix}, \label{2f6}
\end{equation}

\begin{equation}
\widehat{\mathbf L}_{{\rm y}4} = \begin{pmatrix} 
-\cos \beta_1 & 0 & 0 & 0 & 0 & -\sin \beta_1 \\
0 & 1  & 0 & 0 & 0 & 0 \\
0 & 0 & \sin \beta_3 & -\cos \beta_3 & 0 & 0 \\
0 & 0 & \cos \beta_3 & \sin \beta_3 & 0 & 0 \\ 
0 & 0 & 0 & 0 & 1 & 0 \\
-\sin \beta_1 & 0 & 0 & 0 & 0 & \cos \beta_1 
\end{pmatrix}. \label{2f7}
\end{equation}

Even though we have managed to express non-unitary operators as LCUs,
the actual realization of the ensuing algorithm in a quantum computer needs some thought.
In particular one is forced into an embedding into a higher dimensional unitary matrix from which measurements
are required to recover non-deterministically the probability of success \cite{childs2012}.  
We do not pursue this further here
since currently there is no quantum computer currently available on which to perform an LCU-QLA
simulation.

In the meantime, we have implemented \eqref{2e14} on ${\it Perlmutter}$ -- a
classical supercomputer at the National Energy Research Scientific Computing Center.
\section{Simulation of transient scattering by a localized dielectric medium}
\label{sec:5}
\subsection{QLA Code Performance on ${\it Perlmutter}$ :  strong and weak scaling up to 110,000 cores}
The QLA framework is naturally suited for parallel execution.  The unitary collision operators and the two
non-unitary potential operators entangle on-site qubits only at each lattice node.  The streaming operator is a
simple shift operation for internal lattice points in each processor domain.  However for boundary lattice
points synchronized nearest-neighbor communications through ghost-cell updates is required.  This locality
minimizes communication overhead and avoids costly global operations, thereby enabling excellent 
scalability on distributed-memory architectures.

Even though the results reported here are for 2D spatial variations, the performance for 3D spatial variations is
more demanding and the scaling results will be reported for 3D runs.  Since QLA is so modular, it is relatively
straightforward to extend the 2D code to 3D by including MPI instructions for boundary points of the processors.

In Table 1, we report the strong scaling of our 3D QLA Maxwell code.  In strong scaling one fixes the grid and increases
the number of cores.  Ideally, the wall-clock time should decrease, proportional to the (number of cores)$^{-1}$.
\begin{table}[H]
\begin{center}
\begin{tabular}{|c|c|c|c||} \hline 
{\bf STRONG} & cores & {\bf wall-clock} [ideal] (s) & efficiency \% \\ \hline \hline
{\it grid $4800^3$} & 32,768  & {\bf 307.1} & 100  \\  \hline
 & 65,536 & {\bf 150.0} [153.6] & 102.4   \\  \hline
 & 110,592 & {\bf 87.8} [91.0]  & 103.6 \\  \hline
\end{tabular}
\end{center}
\caption{{\bf Strong  scaling of the 3D MPI QLA Maxwell code on \it{PERLMUTTER.}
Here one fixes the grid and increases the number of cores.  For ideal parallelization
the wall clock time should scale as (number of cores)$^{-1}$.
 The 3D QLA  code shows slight superlinear scaling, on using the $32,768$-core baseline.
 } }
\end{table}
The superlinear behavior is likely due to improved cache utilization, more efficient memory access, and reduced 
memory  contention as the workload per core decreases. Overall performance is further supported by the strictly 
local communication pattern, which involves only nearest-neighbor exchanges.

In weak scaling, the computational workload per core is held constant as the total number of cores increases. For example, 
increasing the core count by $2^3$ corresponds to doubling the side length of the 3D computational domain. 
As shown in  Table 
2, the wall-clock time varies by less than $0.4\%$ as the simulation scales from 512 to 110592 cores, indicating a very well 
parallelized code.
%In weak scaling one fixes the work done by each core as one increases the number of cores used.  For example, if 
%one increases the number of cores by $2^3$ then one would increase the 3D cube length by a factor of $2$.
%From Table 2 one sees less than $0.4\%$ variations in wall clock time as one scales from $512$ cores to $110,592$
%cores, indicating a very well parallelized code.
\begin{table}[H]
\begin{center}
\begin{tabular}{|c|c|c|c|} \hline 
{\bf WEAK} & Grid    &  cores & {\bf wall-clock} (s) \\ \hline \hline
& $800^3$  & 512  & {\bf 87.51}  \\  \hline
& $1200^3$  & 1,728 & {\bf 87.92}  \\  \hline
& $1600^3$  & 4,096  & {\bf 87.90}    \\  \hline
& $2400^3  $  & 13,824 & {\bf 87.90}  \\ \hline
& $4800^3  $  & 111,592 & {\bf 87.83}  \\ \hline
\end{tabular}
\end{center}
\caption{{\bf Weak scaling of 3D QLA code on \it{PERLMUTTER}} \\
}
\end{table}
This behavior reflects the locality of the algorithm, in which communication is confined to nearest-neighbor exchanges, allowing the method to maintain high efficiency even at very large core counts.

\subsection{QLA 2D Initial Conditions}

We now discuss some QLA simulation results from the evolution equation \eqref{2e14} for 2D scattering from an
isotropic dielectric medium $n(x,y)$ of an incident 1D Gaussian wave packet with electric field ${\bf E}(y, t =0) =
{\bf \hat{z}} E_0 Exp(-y^2/b^2) Cos(2 \pi y/ \lambda) $ with the corresponding $B_x(y,t = 0) {\bf \hat{x}}$ magnetic field.  
The plane of incidence is the $x-y$ plane.  In the QLA simulations, the 2D lattice grid is $16384^2$ and the
elliptical dielectric is $1600 \times 716$.  The incident pulse wavelength $\lambda = 100$ with pulse thickness
$600, \, (b=44.7)$.   These parameters have the scattering in the Mie range \cite{hulst, bohren, kong}.
There is a thin continuous boundary layer region, $40 \times 18$ at the elliptical surface in which the dielectric 
refractive index rises from the vacuum $n_v=1$ to its uniform interior of $n_d=3$.
We will also do simulations for the reverse situation:  a vacuum elliptical bubble embedded in a uniform dielectric.

QLA is an initial value code and no internal boundary conditions are imposed at any time in the scattering
simulation.  The thin continuous boundary layer around the ellipse permit the imposition of Maxwell's equations,
\eqref{3.3} and \eqref{3.4}.  The two divergence equations, \eqref{3.1} and \eqref{3.2}, and the instantaneous
total electromagnetic energy, \eqref{eqA13}, are monitored throughout the QLA simulation.  Both divergence
equations are satisfied to within machine accuracy while the energy conservation is satisfied to seven significant 
figures.  This very slight loss in energy conservation is due to the 2 non-unitary potential operators 
 $\widehat{\mathbf V}_{\rm x}(\delta)$  and  $\widehat{\mathbf V}_{\rm y}(\delta) $ which break the unitarity of the 
 18 operator sequence in \eqref{2e14}.
 
 \subsection{Scattering off an elliptic dielectric within a vacuum}
 
 In Figs 2-13 we show the time evolution of the electric field $E_z$ as the wave packet propagates in the vacuum 
 and scatters from the localized dielectric.  The initial approach of the wave packets towards the dielectric is shown
 in FIg. 2.  We refer to that part of the dielectric facing the incident wave packet as the front end of the dielectric;
 and the back end being the opposite boundary.  Slightly later in time, one sees the 
 beginning of the slowing down of the wave
 packet inside the dielectric since its group velocity is reduced because of its higher dielectric refractive index 
 $n_d=3 > n_v=1$, Fig. 3.
 
 In Fig. 4, the front portion of the wave packets in the vacuum has moved just beyond the back end of the dielectric,
 while the fields inside the dielectric are lagging by a factor of $n_v/n_d$.  As the wave packet propagates past the
 back end, a portion of the wave packet in the vicinity of the dielectric develops filamentary structures.  This is 
  evident in Figs. 5 and 6.  Such structures also occur in the frequency domain when electromagnetic waves
  in a plasma scatter off a circular cylindrical plasma \cite{ram2016,ram2013,ioann2017}.  The primary
reason for these filaments is topological. The phase front of
the wave packet is planar while the dielectric is elliptical.
This requires the wave to curve around the dielectric while
satisfying boundary conditions on the surface of the dielectric
 continuity of the tangential component of the electric field
and continuity of the normal component of the magnetic field.
Essentially, it is a mapping between Cartesian and elliptical
coordinates.
 
The wave fields inside the dielectric reach the back end and
are partially transmitted into vacuum and partially reflected
within the dielectric. Figures 7, 8, and 9 show the progression
of the transmitted and reflected waves. The striations in the
field structures are boundary effects. Further along in time,
Fig. 7 reveals the decoupling of the wave packet from the fields
inside the dielectric. Subsequently, the radiation patterns are
determined by the fields inside the dielectric; in effect, a leaky
dielectric filled wave guide being an antenna. The angular
variation of the field around the z-axis leads to striations inside
the dielectric. Later in time, as the fields from the back end
of the dielectric reach the front end, there is transmission into
vacuum. Figures 10 and 11 show the time evolution of the
reflected fields as they propagate in a direction opposite to that
of the original wave packet. In contrast to side scattering of the
incident wave packet in Figs. 4 - 7, the secondary emission
of the fields from the dielectric leads to both side scattering
and backscattering, as displayed in Fig. 12. It is noteworthy
that backscattering did not occur during the initial transition
of the wave packet through the dielectric; rather, it was due
to the wave fields that were left behind and trapped within
the dielectric. Initially, there was side scattering followed by
forward scattering of the wave packet. The backscattering is
a secondary effect. The fields inside the dielectric continue
to bounce around and lead to further transmission through
the boundary, albeit, with lesser intensity. Figure 13 displays
scattered fields from multiple bounces of the wave fields
partially trapped within the dielectric. From these simulations
we note that an observer outside the dielectric will notice
several bursts of electromagnetic waves emanating from the
dielectric well after the initial wave packet has propagated far
away from the dielectric.

\begin{figure}[!t]
\centerline{\includegraphics[width=0.8\columnwidth]{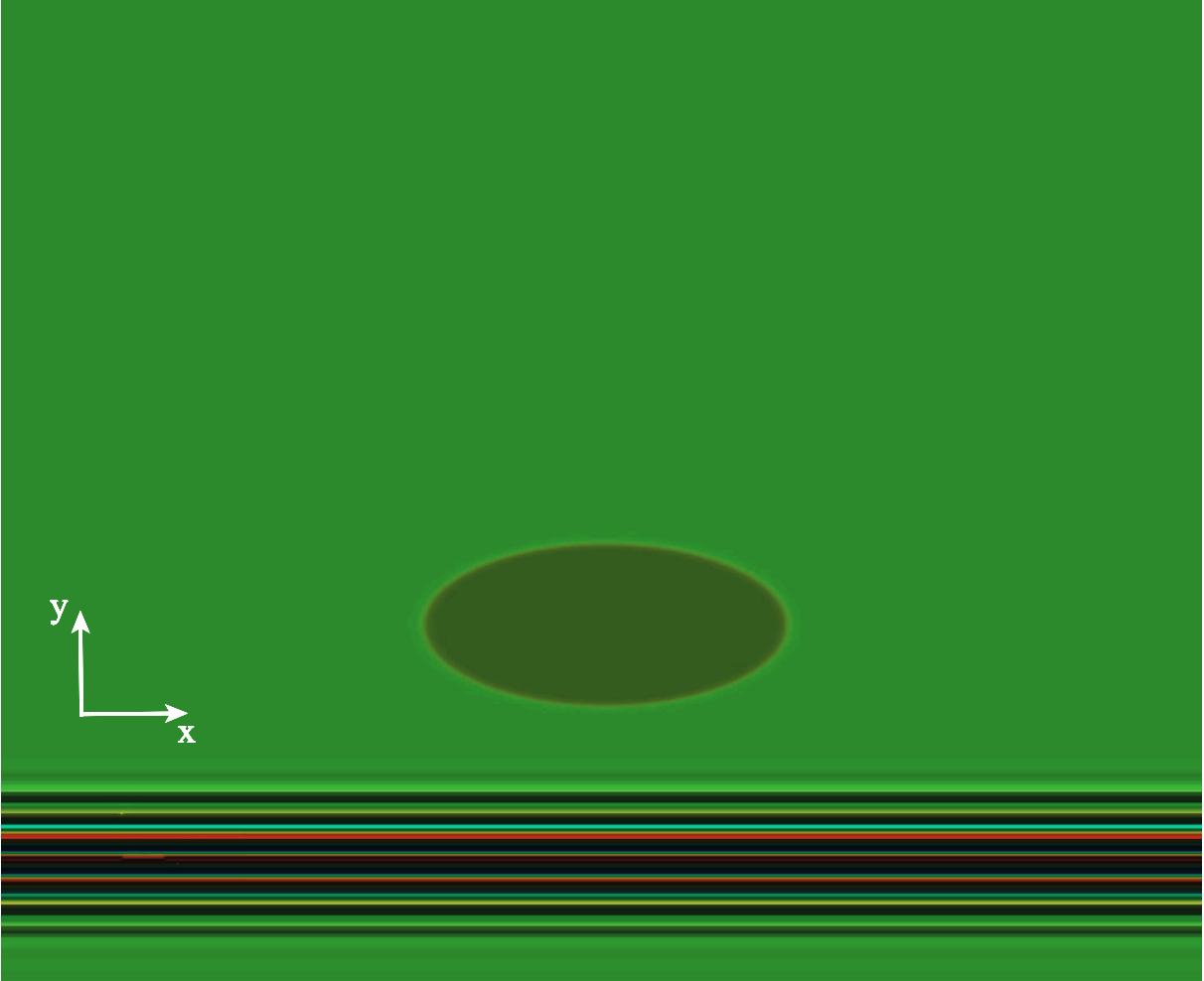}}
\caption{The initial  electric field $E_z(y,t=0)$ component of the 1D Gaussian wave packet as it propagates in a vacuum
in the $y$-direction towards an elliptical dielectric.  The colors of the wave packet indicate the field strengths, with
the maximum amplitude at the central-$y$ value.}
\label{fig2}
\end{figure}

\begin{figure}[!t]
\centerline{\includegraphics[width=0.8\columnwidth]{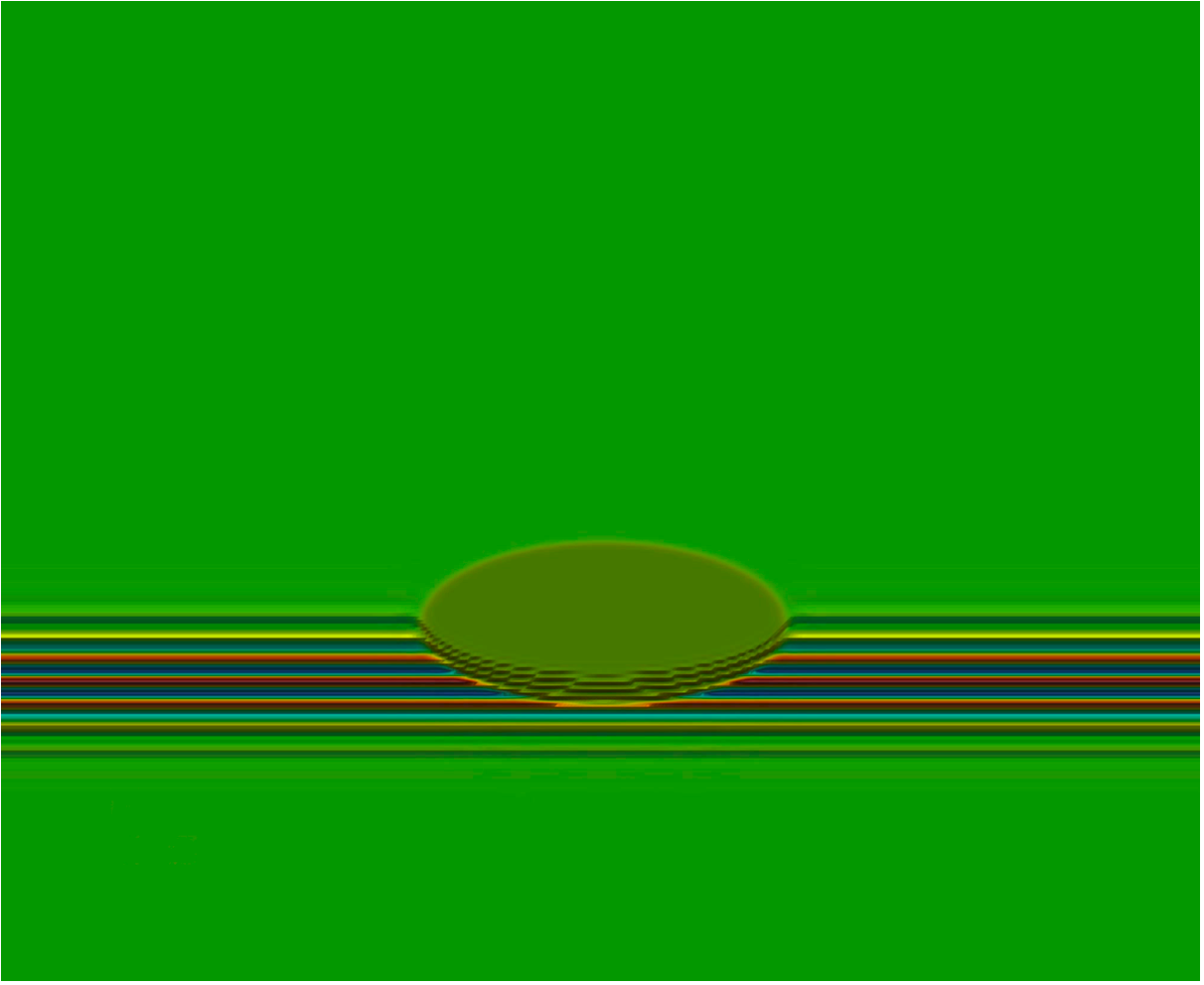}}
\caption{As the wave packet begins its interaction with the dielectric, its phase velocity inside the dielectric slows down. 
The planar wave front of the wave packet gets distorted by the curved surface of the dielectric.}
\label{fig3}
\end{figure}

\begin{figure}[!t]
\centerline{\includegraphics[width=0.8 \columnwidth]{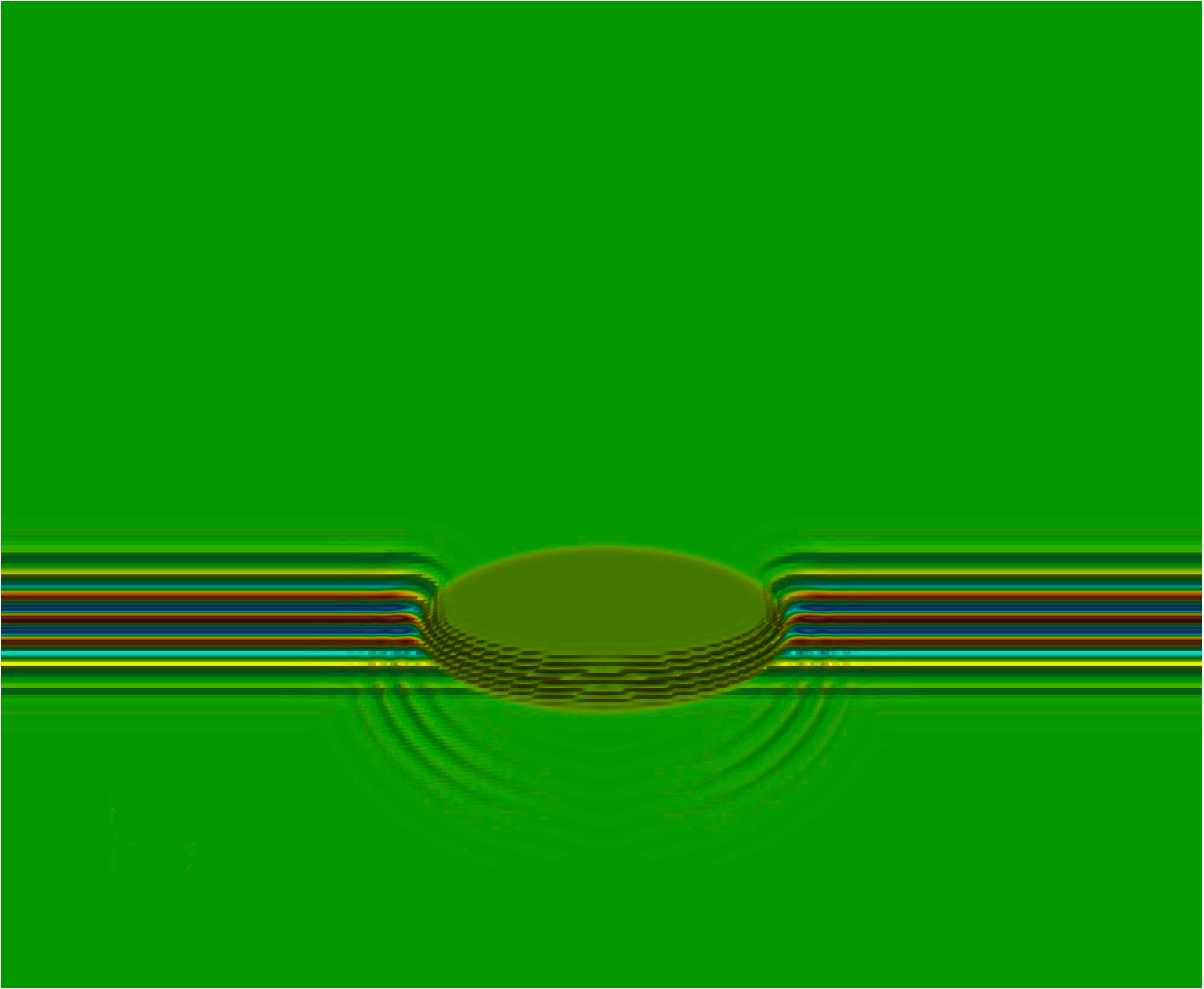}}
\caption{There appear wave patterns of side-scattering as the wave packet continues to propagate past the dielectric. The
lagging fields inside the dielectric form their own spatial structure.}
\label{fig4}
\end{figure}

\begin{figure}[!t]
\centerline{\includegraphics[width=0.8 \columnwidth]{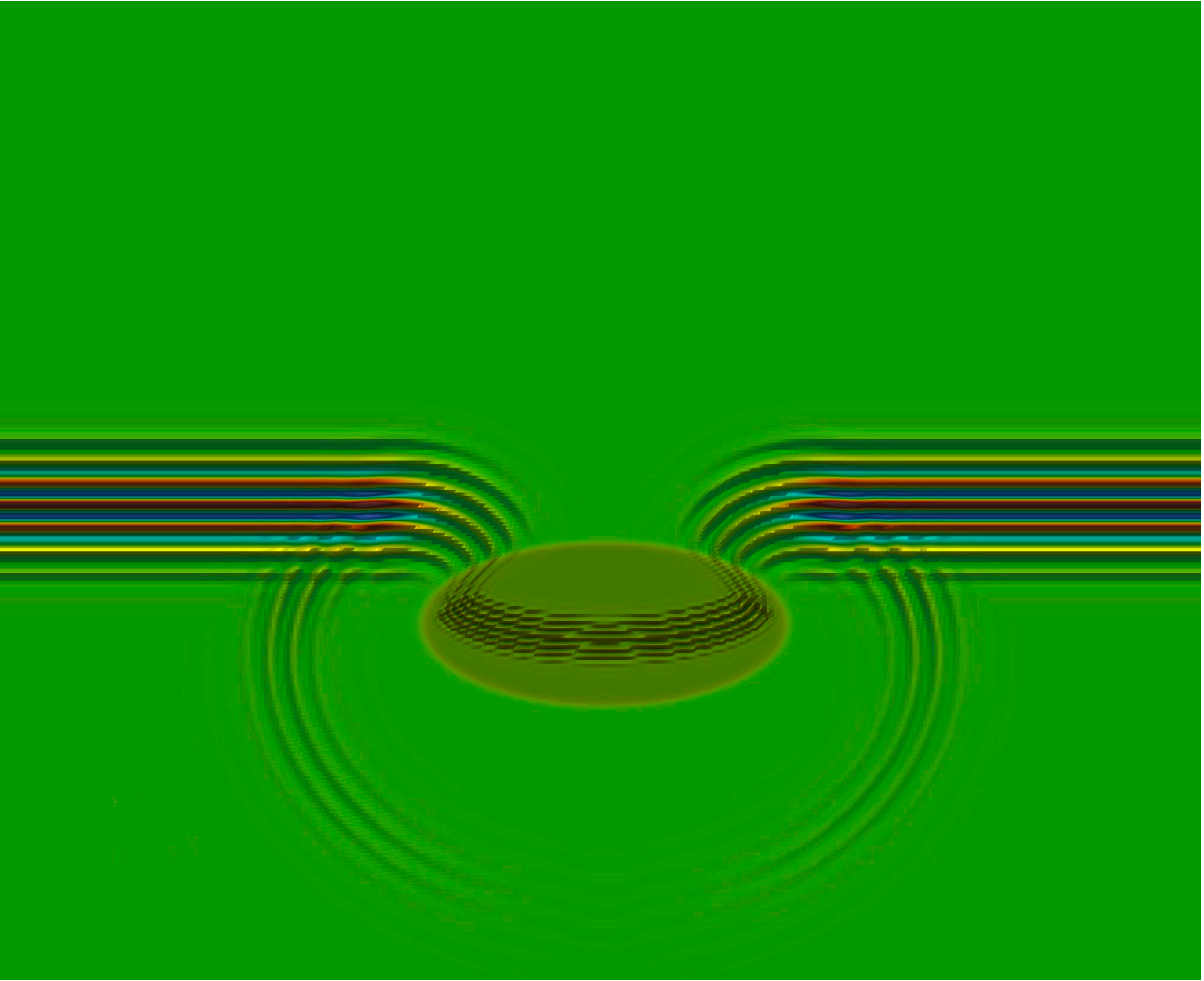}}
\caption{The appearance of filamentary structures in the wave pattern of the wave packet, while fields inside the
dielectric propagate from the front end towards the back end.}
\label{fig5}
\end{figure}

\begin{figure}[!t]
\centerline{\includegraphics[width=0.8 \columnwidth]{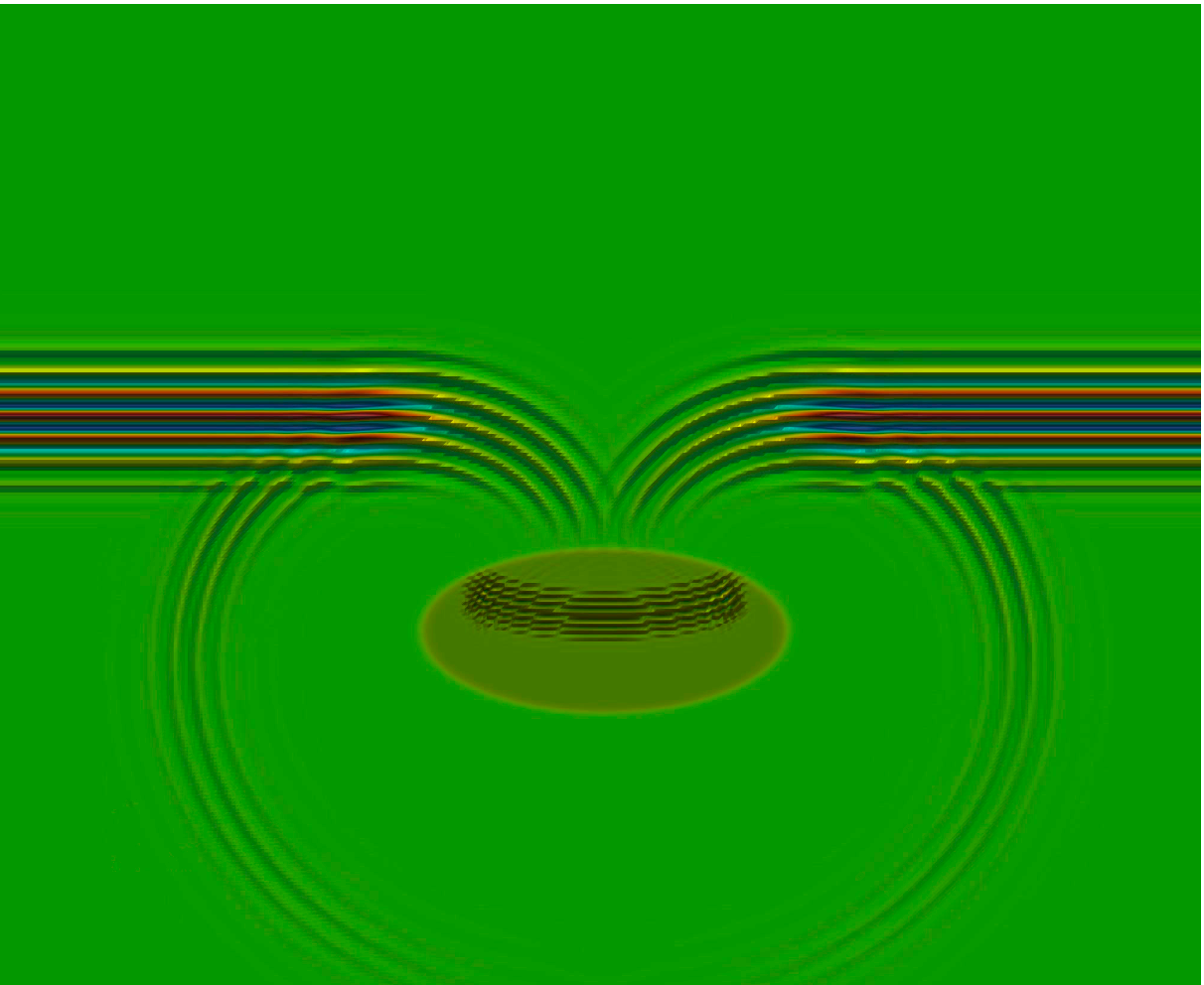}}
\caption{The emerging wave patterns as the simulations advance in time. The side-scattered fields, generated earlier,
  continue to propagate away from the elliptical dielectric. The fields outside the dielectric are beginning to decouple
from the fields inside the dielectric.}
\label{fig6}
\end{figure}

\begin{figure}[!t]
\centerline{\includegraphics[width=0.8 \columnwidth]{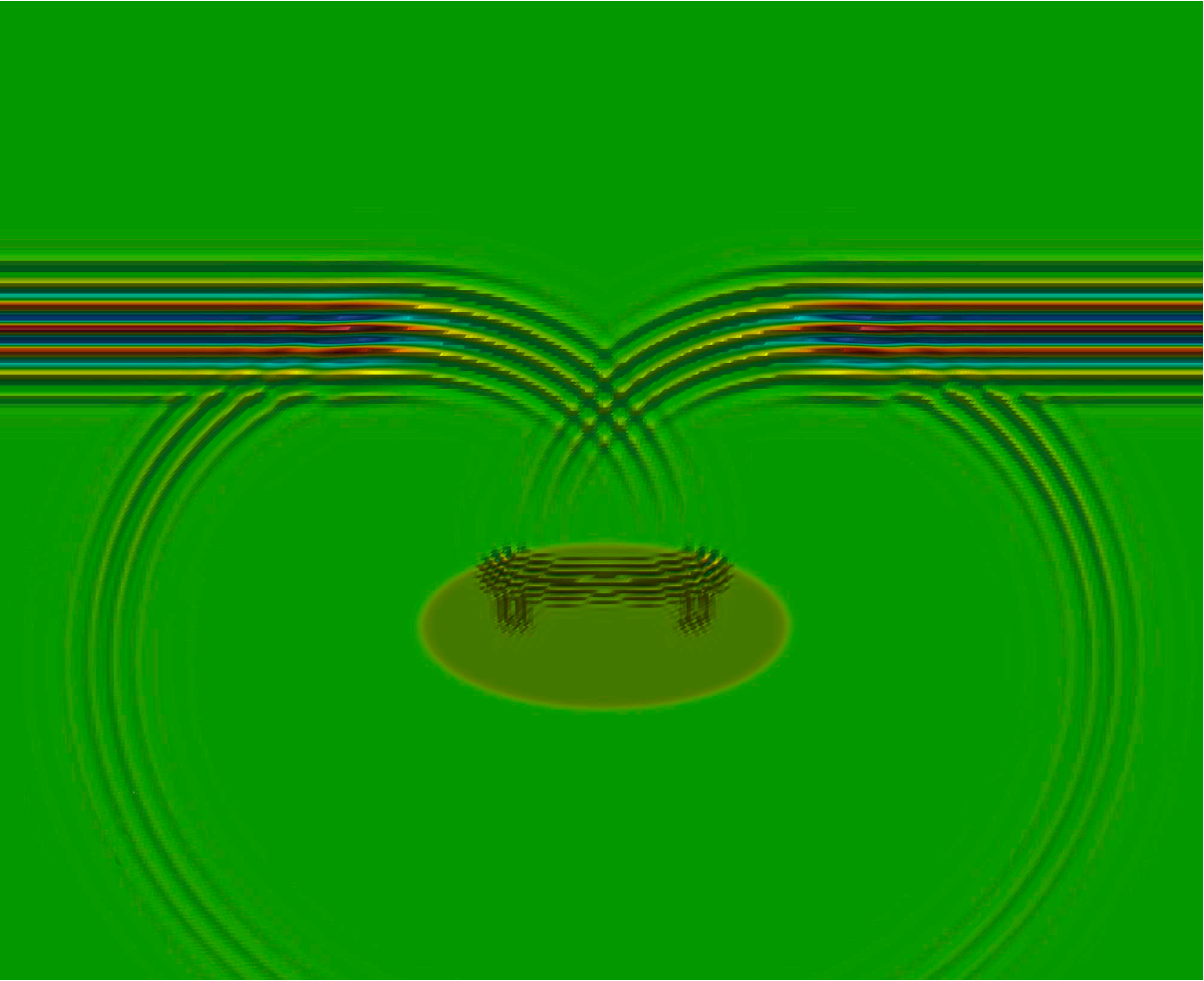}}
\caption{The wave fields inside are getting reflected from the back end of the dielectric.
The fields inside and outside of the dielectric are essentially decoupled. The striations in fields inside are
a result of the curved boundary.}
\label{fig7}
\end{figure}

\begin{figure}[!t]
\centerline{\includegraphics[width=0.8 \columnwidth]{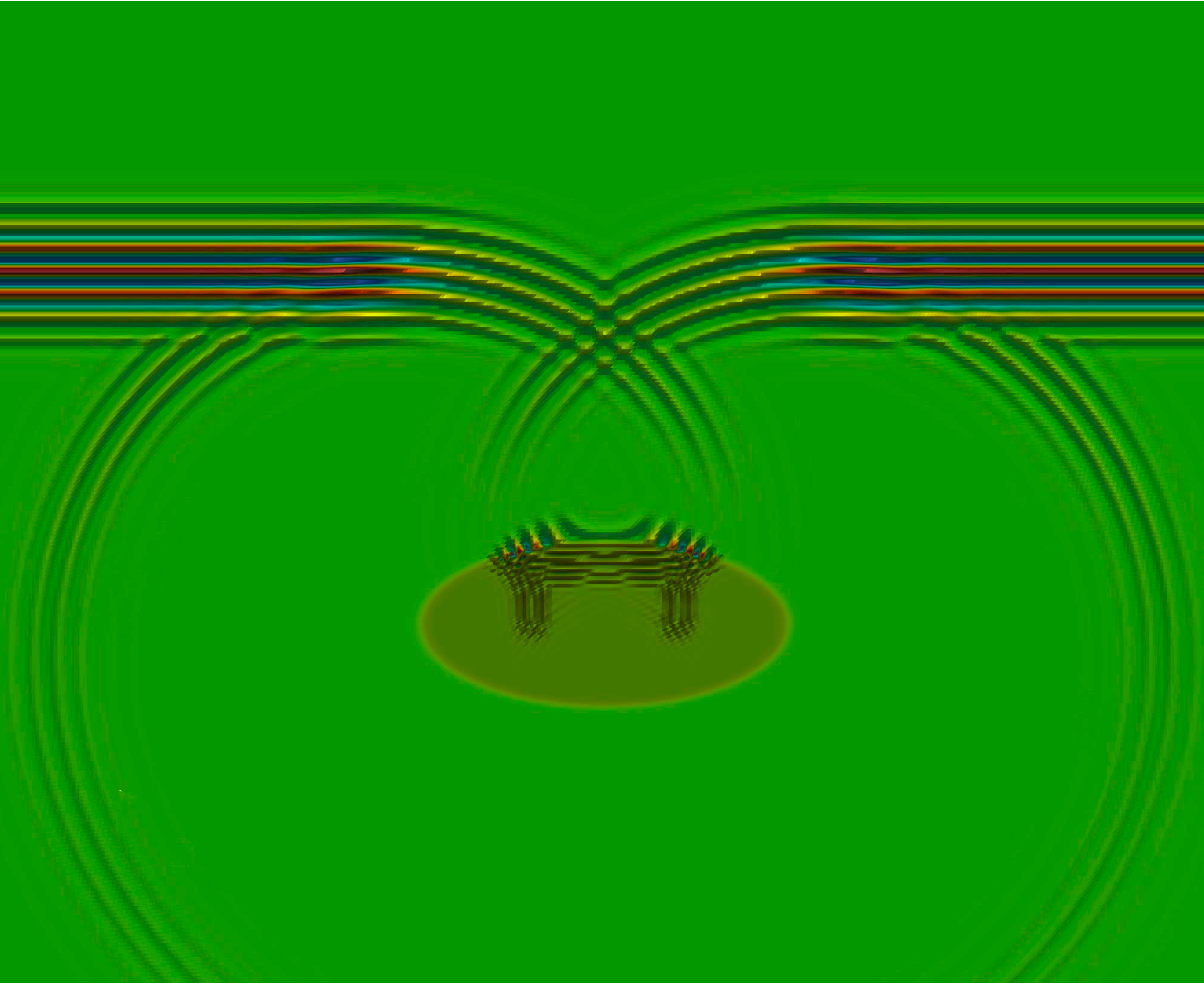}}
\caption{The forward scattered wave patterns as the wave packet separates from the dielectric.
We also note that there is transmission of the fields from inside the dielectric.}
\label{fig8}
\end{figure}

\begin{figure}[!t]
\centerline{\includegraphics[width=0.8 \columnwidth]{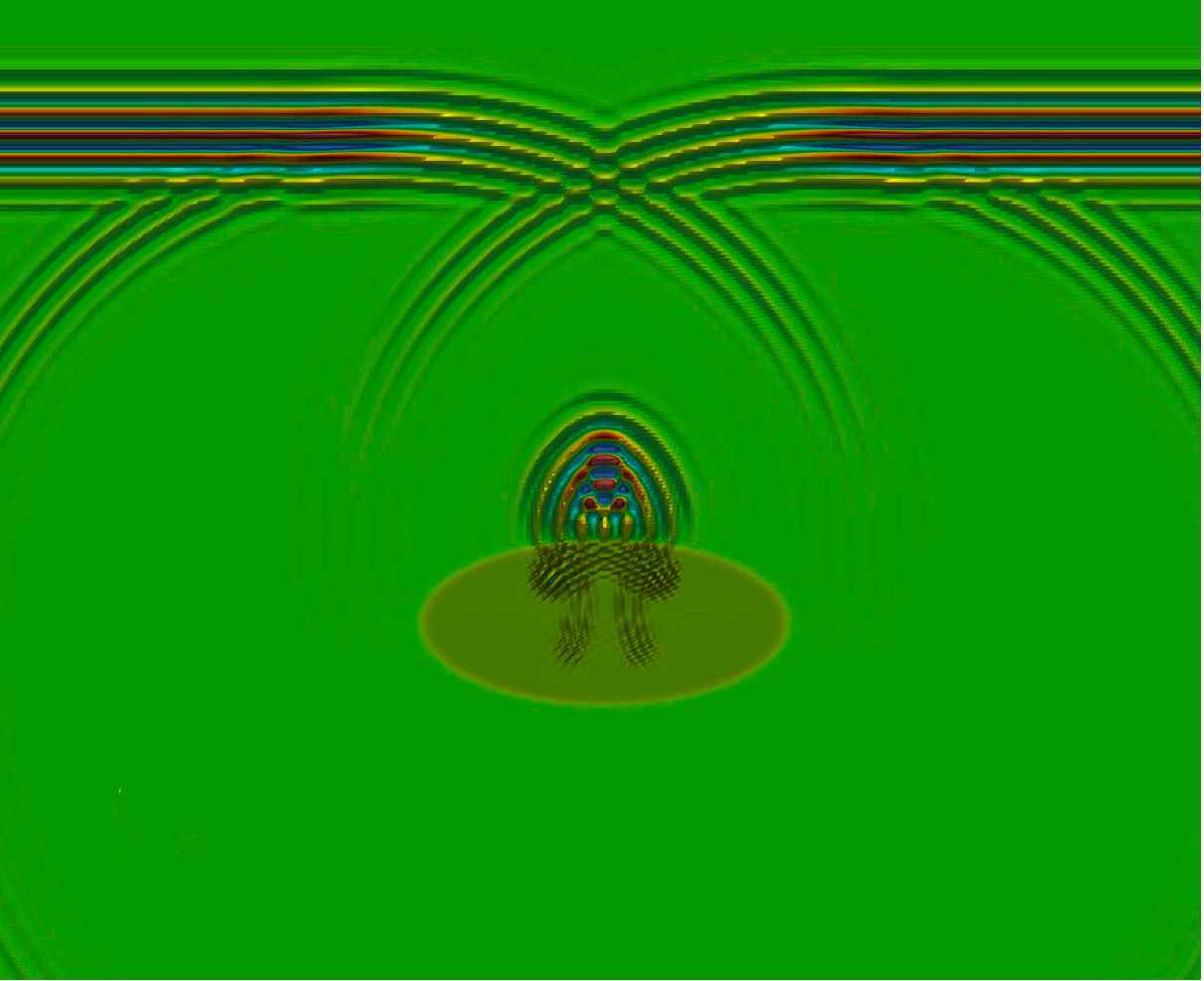}}
\caption{The secondary forward scattering due to fields inside the dielectric is evident.
The fields reflected from the back end are reaching the front end of the dielectric.}
\label{fig9}
\end{figure}

\begin{figure}[!t]
\centerline{\includegraphics[width=0.8 \columnwidth]{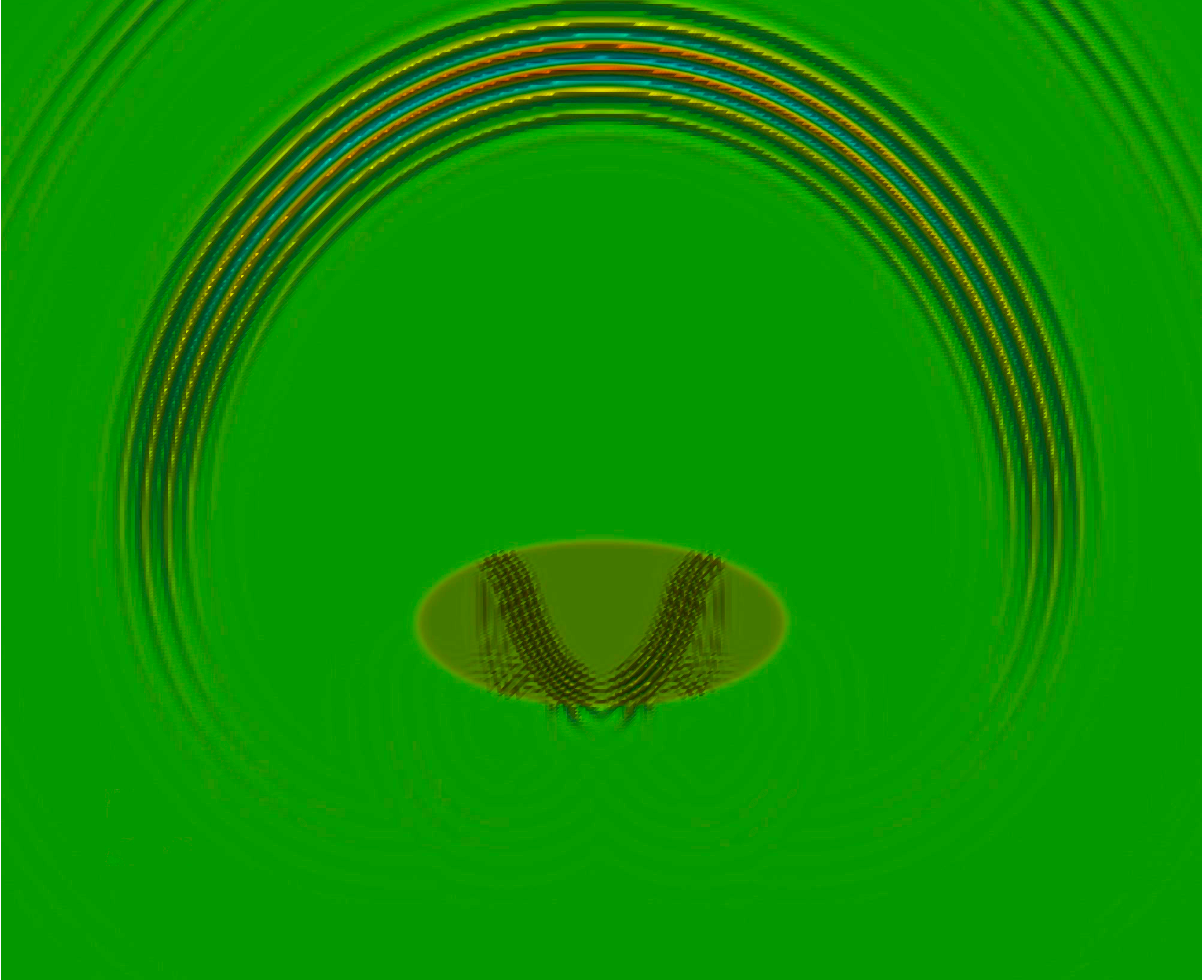}}
\caption{The secondary forward scattered fields propagate away from the dielectric, while
a fraction of the internally reflected fields are being transmitted through the front end. 
This is the beginning of back scattered fields.}
\label{fig10}
\end{figure}

\begin{figure}[!t]
\centerline{\includegraphics[width=0.8 \columnwidth]{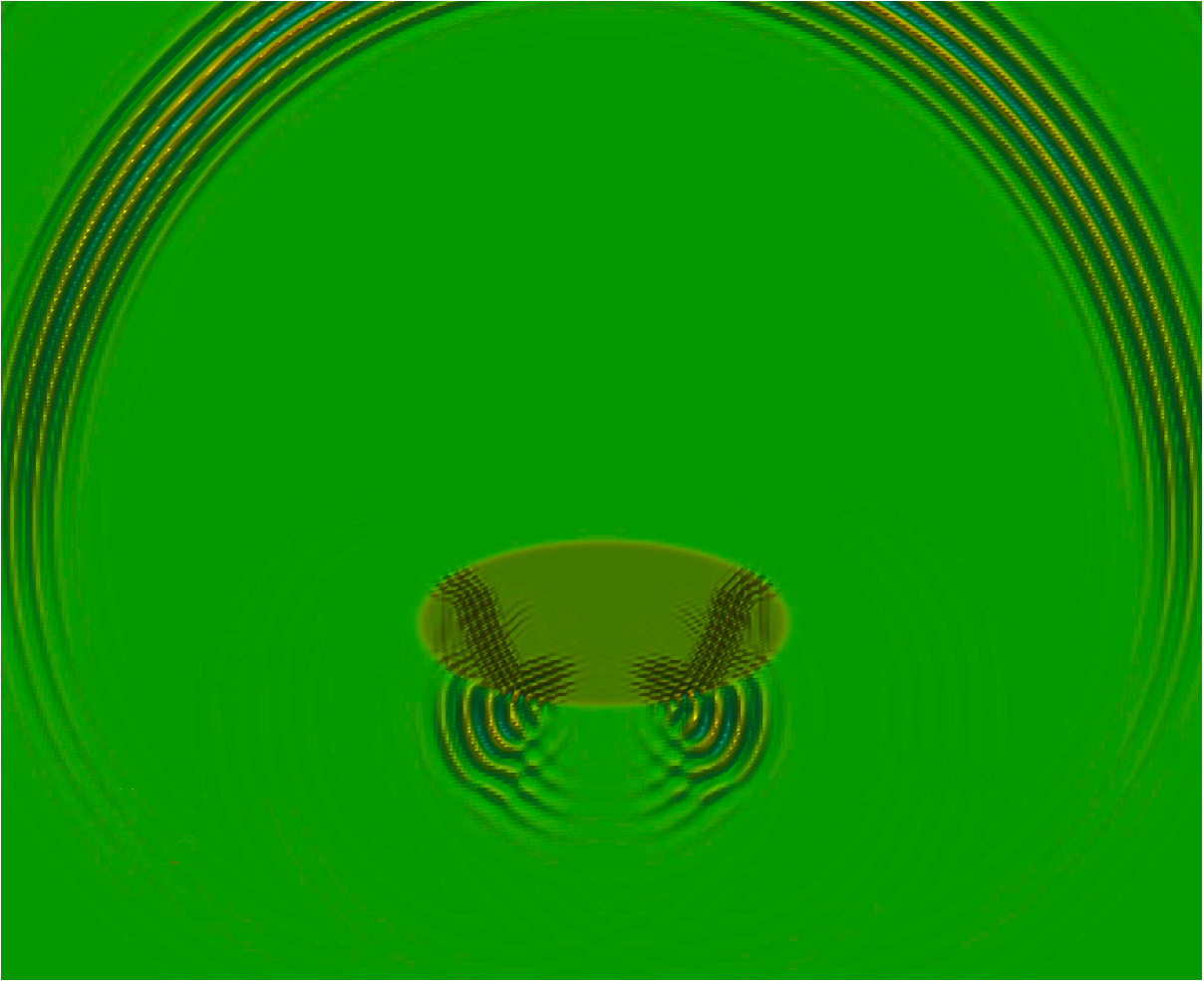}}
\caption{The internal fields are propagating along the major axis of the ellipse, while
back scattered fields are more discernible and propagating away from the dielectric.}
\label{fig11}
\end{figure}

\begin{figure}[!t]
\centerline{\includegraphics[width=0.8 \columnwidth]{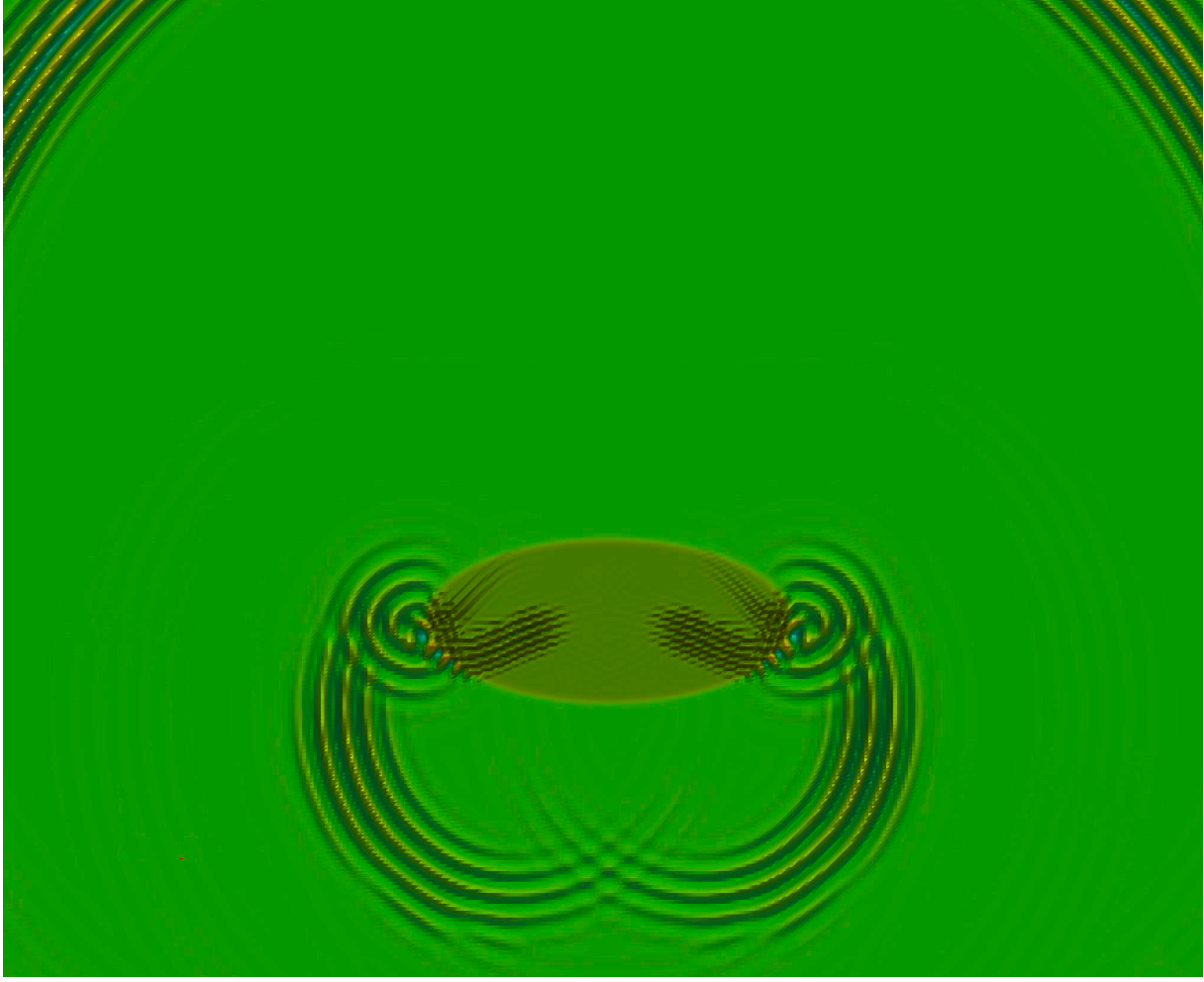}}
\caption{The internal wave fields which were propagating along the major axis of the ellipse lead to
secondary side scattering as well as back scattering.}
\label{fig12}
\end{figure}

\begin{figure}[!t]
\centerline{\includegraphics[width=0.8 \columnwidth]{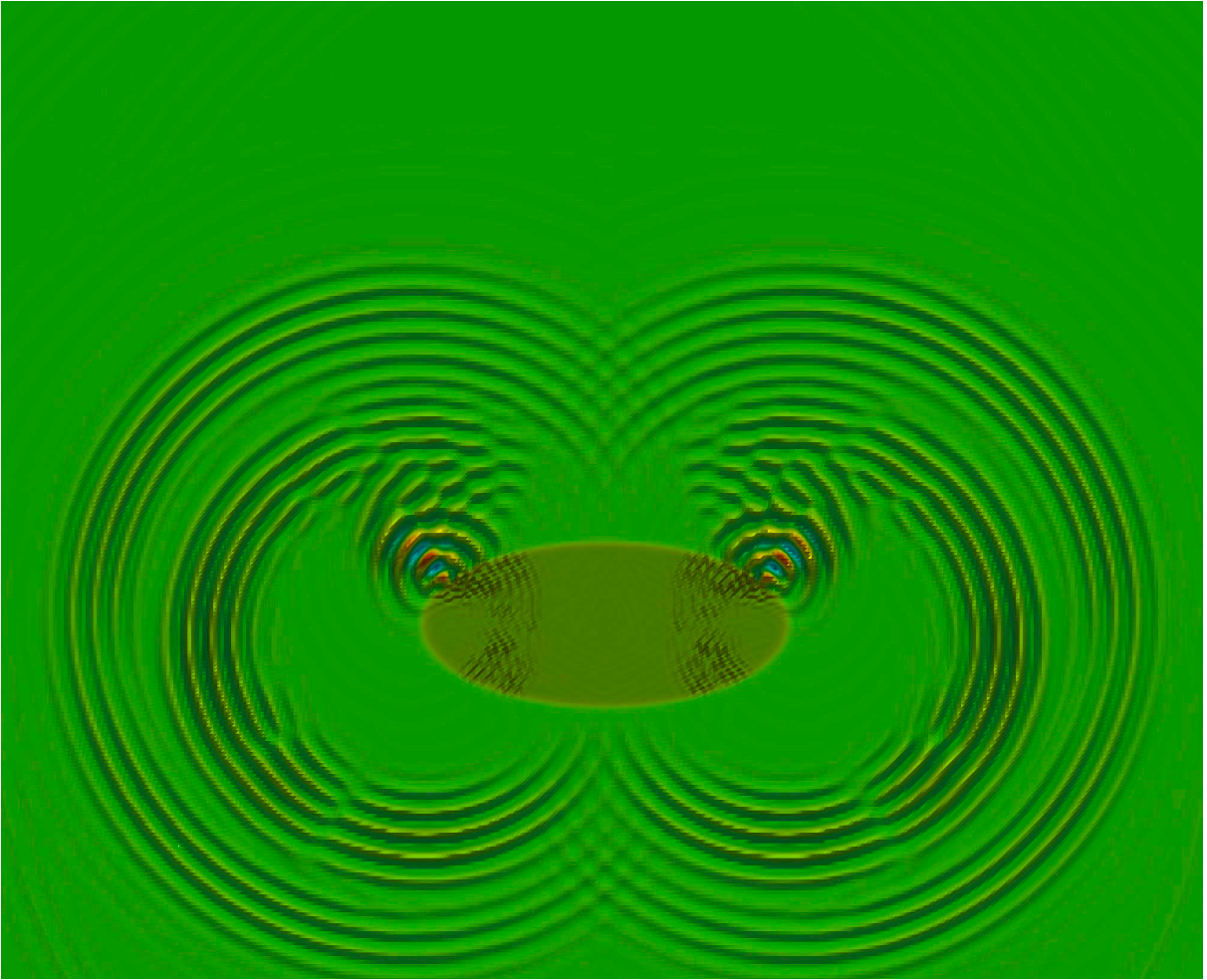}}
\caption{The two distinct backscattered wave patterns indicate at least two transmissions at two different times as
fields inside the dielectric reflect back-and-forth from the front and back boundaries.}
\label{fig13}
\end{figure}

\subsection{Magnetic fields}
The initial wave packet in Fig. 1, is propagating along the
y-axis with its electric field vector along z,  perpendicular
to the plane of incidence, and the magnetic field vector is along $x$.
Since there is uniformity along z, and the electric field
is continuous across the surface of the dielectric, 
$\nabla \pmb{\cdot}  {\mathbf D} \left( {\mathbf r}, t \right) \,= \ 0,$ %\label{3.1} 
 is satisfied throughout the entire simulation without any
other Cartesian components contributing to the electric field
vector. This is not the case for the magnetic field. As there are
no surface currents induced at the boundary of the dielectric,
the tangential component of the magnetic field has to be
continuous. The initial magnetic field of the wave packet,
polarized along x, is not tangential along the entire surface
of the elliptic dielectric. In order to comply with the divergence condition
$\nabla \pmb{\cdot}  {\mathbf B} \left( {\mathbf r}, t \right) \,= \ 0,$ %\label{3.2} 
 the boundary condition requires the generation of a $B_y$  component of
the magnetic field. These standard arguments of standard electromagnetic theory
are completely upheld by our QLA simulations in which no internal boundary conditions
are implemented.

Initially, the only component of the magnetic field that is non-zero is $B_x(y,0)$ and from
Maxwell equations it is proportional to the initial $E_z(y,0)$.  As the wave packet starts its
interaction with the dielectric, the effect on the $\mathbf{E}$ is the relatively small distortion 
in $E_z(x,y)$ as shown in Fig. 3 with a similar distortion in $B_x(x,y)$, Fig. 15, but the QLA
self-consistently creates a small localized non-zero $B_y(x,y)$, FIg. 16, so that for all times 
$\nabla \pmb{\cdot}  {\mathbf B} \left( x,y, t \right) =  0$ to machine accuracy -- even though
the QLA is a set of colllide-stream-potential operators that are based on perturbatively
recovering the time evolution subset of Maxwell equations \eqref{3.3} and \eqref{3.4} 
to second order in the lattice mesh size $\delta$.
Figure 17 (for $B_x(x,y)$) and Fig. 18 (for $B_y(x,y)$) are the corresponding magnetic field
components to the electric field $E_z(x,y)$, Fig. 7, at the time instant when the wave packet 
separates from the dielectric.  In FIg. 17 one sees that 1D aspects of the initial wave pulse are
evident far from the dielectric scatterer and these are totally absent in $B_y$ in Fig. 18.

%%%%%%%%%%%%%%%%%%%%%%%%%%%%%%%%%%%%%%%%

\begin{figure}[!t]
\centerline{\includegraphics[width=0.8 \columnwidth]{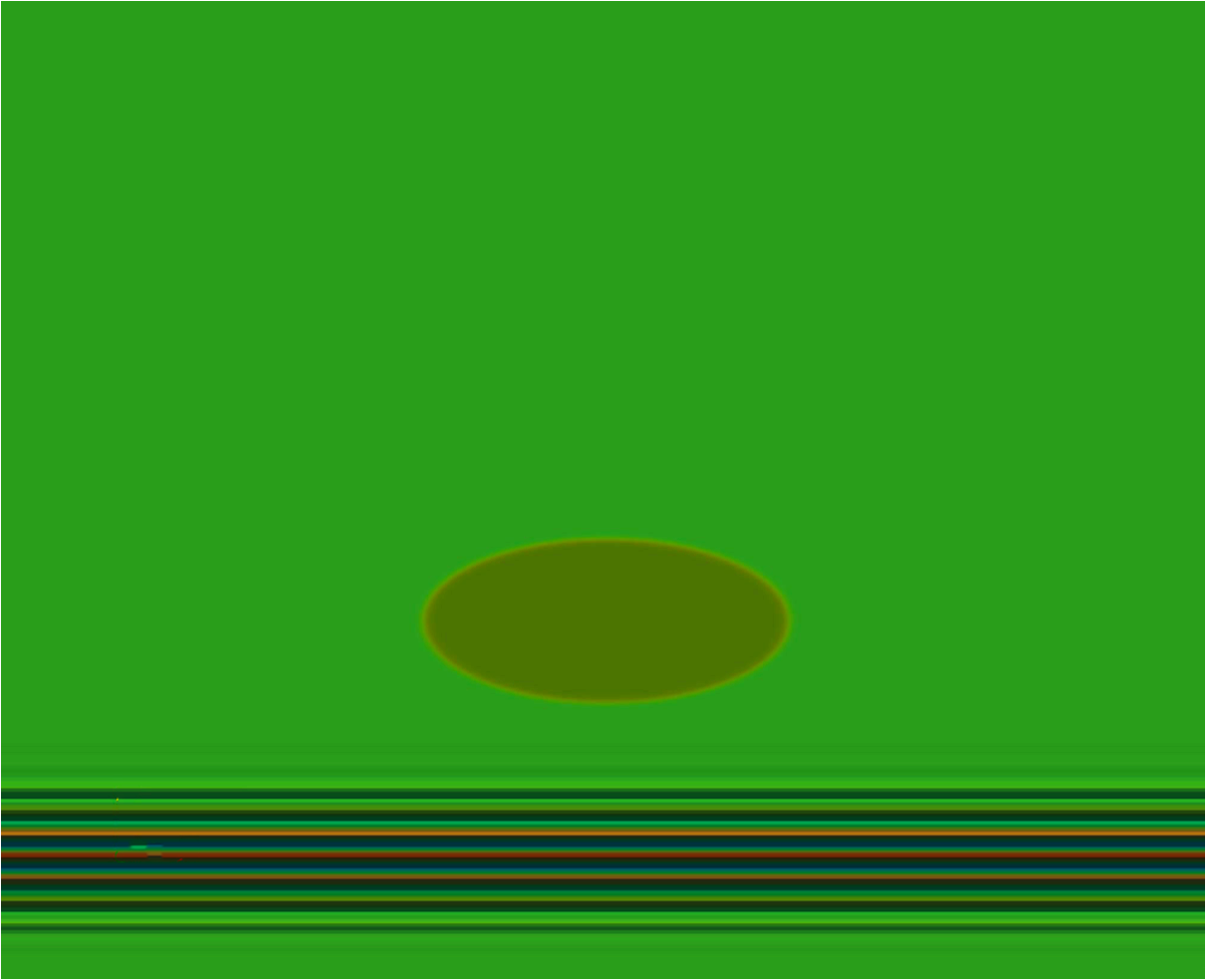}}
\caption{The $x$-component of the magnetic field $B_x$ for the wave packet propagating 
in vacuum and approaching the dielectric. As expected, this figure looks the similar to Fig. \ref{fig2}.}
\label{fig14}
\end{figure}

%\begin{figure}[!t]
%\centerline{\includegraphics[width=0.8 \columnwidth]{Fig14}}
%\caption{$B_y$ at the same instant of time as Fig. \ref{fig13}; initially, $B_y = 0$.}
%\label{fig14}
%\end{figure}

\begin{figure}[!t]
\centerline{\includegraphics[width=0.8 \columnwidth]{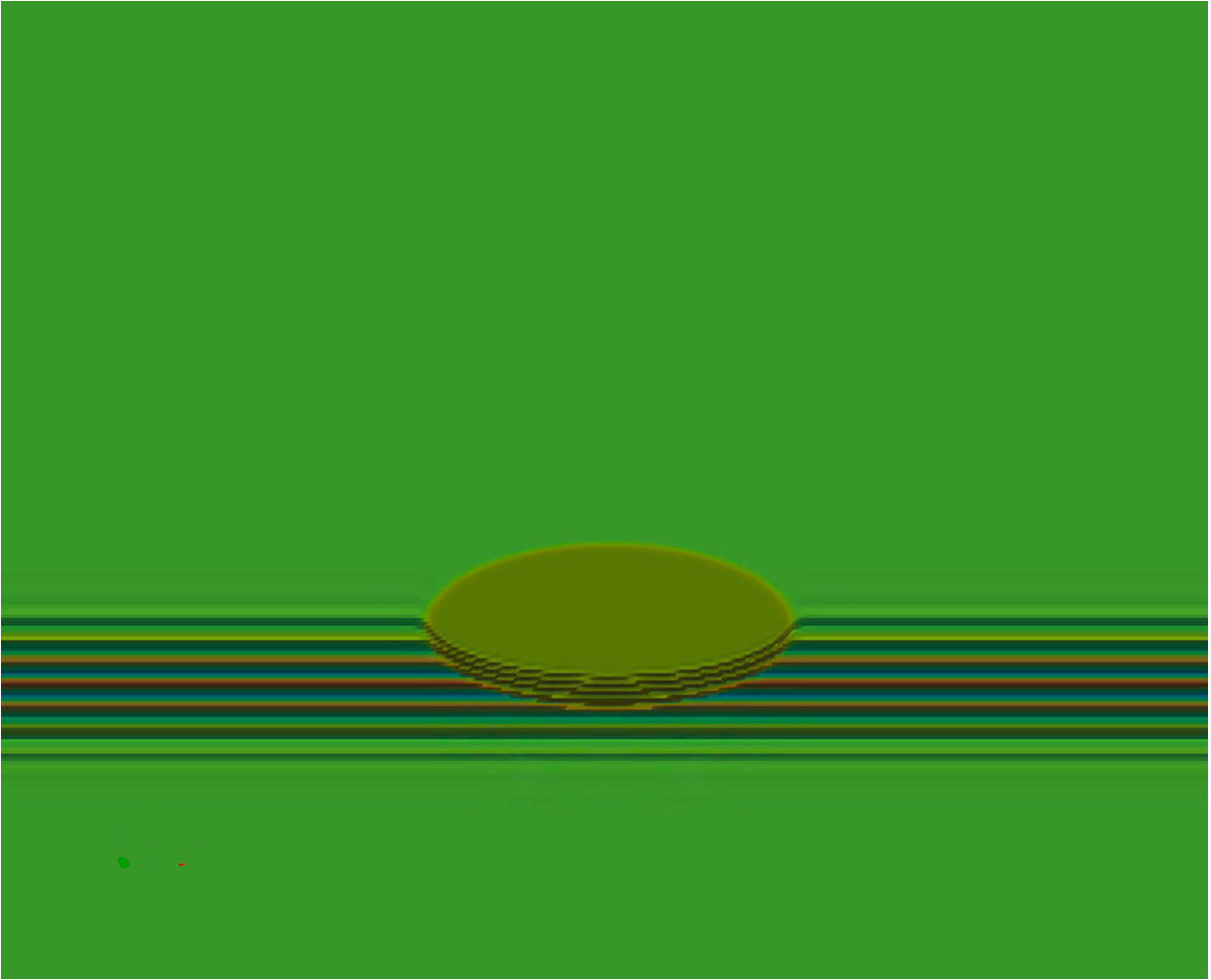}}
\caption{$B_x$ as the wave packet interacts with and propagates past the dielectric. }
\label{fig15}
\end{figure}

\begin{figure}[!t]
\centerline{\includegraphics[width=0.8 \columnwidth]{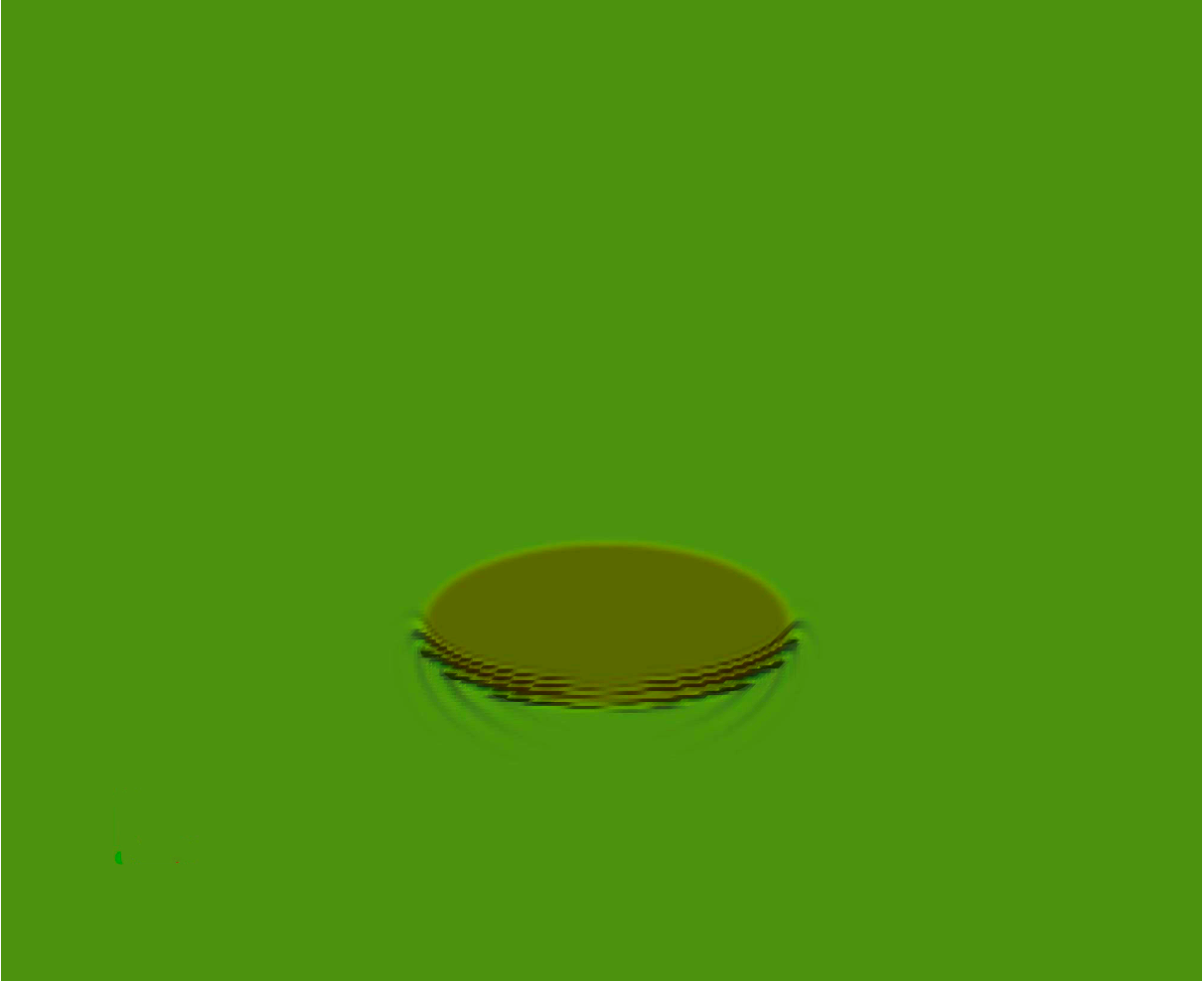}}
\caption{The continuity of tangential component of ${\mathbf B}$ across the surface
of the dielectric induces non-zero $B_y$. The front of the wave packet is planar while
the continuity condition is across an elliptic boundary. The appearance of $B_y$ near
the interface in QLA self-consistently preserves $\nabla \pmb{\cdot}  {\mathbf B} \left( {\mathbf r}, t \right) \,= \ 0$.}
\label{fig16}
\end{figure}

\begin{figure}[!t]
\centerline{\includegraphics[width=0.8 \columnwidth]{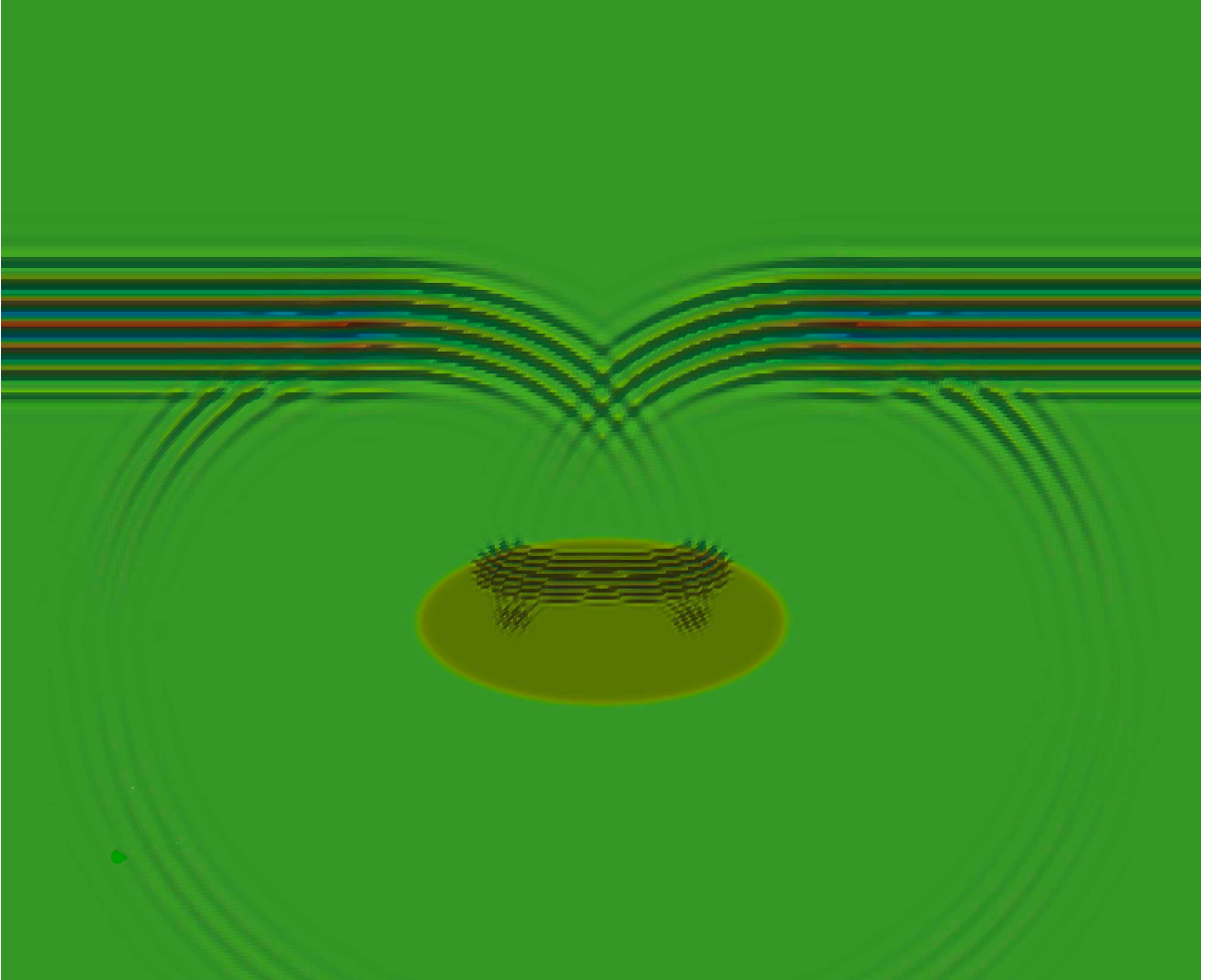}}
\caption{$B_x$ as the wave packet propagates away from the dielectric. The plot
bears resemblance to Fig. \ref{fig6} for $E_z$.}
\label{fig17}
\end{figure}
 
\begin{figure}[!t]
\centerline{\includegraphics[width=0.8 \columnwidth]{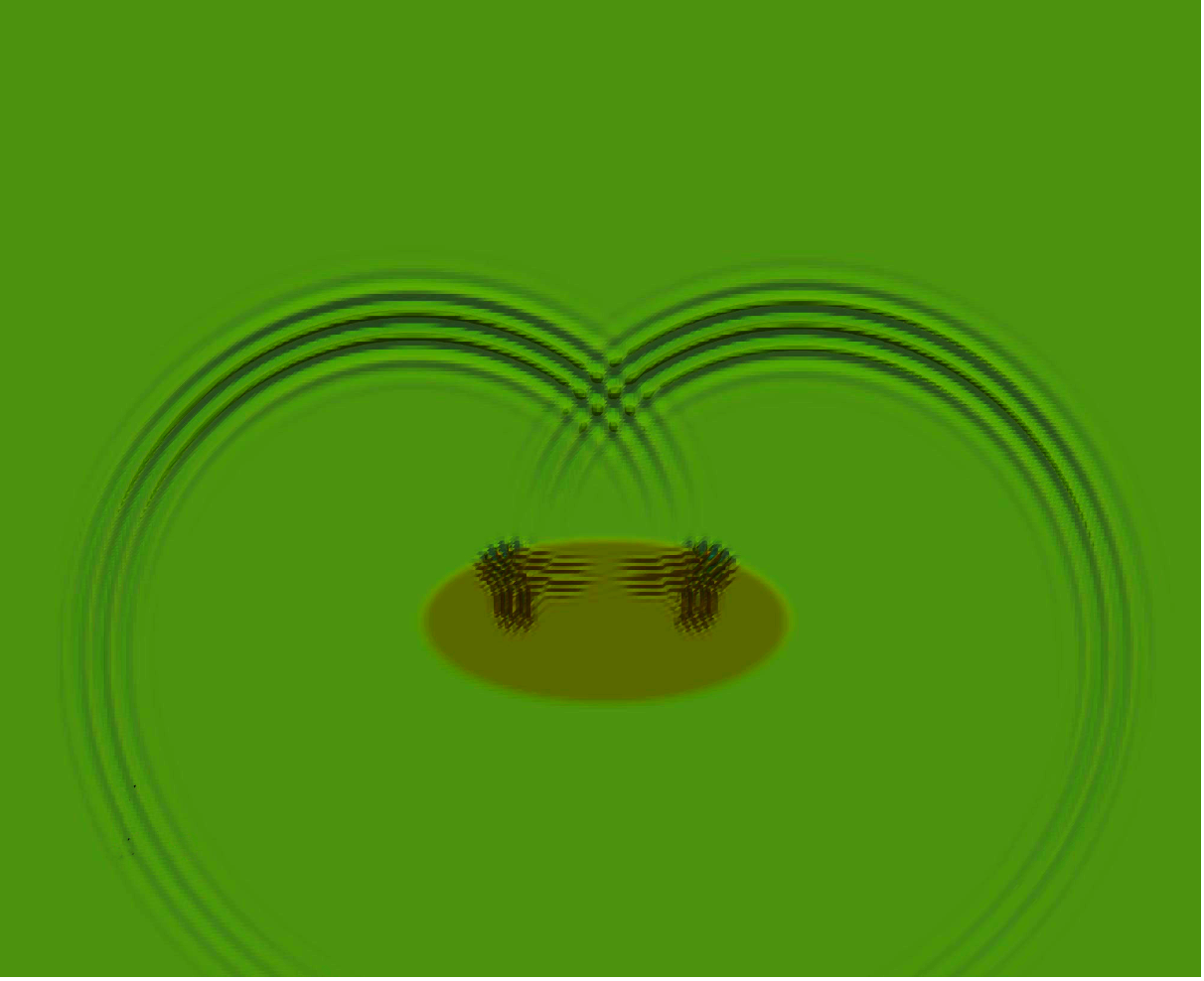}}
\caption{The $y$-component $B_y$ associated with the $B_x$, Fig. \ref{fig17}.}
\label{fig18}
\end{figure}

\subsection{Scattering off an elliptic vacuum bubble embedded in a uniform dielectric medium}
Instead of scattering a wave packet from a elliptic dielectric of larger refractive index , we now
consider the scattering from an elliptic vacuum bubble embedded in a uniform dielectric.
As the wave packet enters the vacuum bubble, its phase velocity increases by a factor of 3 compared 
to that in the dielectric background, Fig. 19.  The reflected fields from the front end lead to side and back
scattering, while the wave fields transmitted through the front
end propagate ahead of the incident wave packet as seen in Fig. 20.

The difference in phase speeds between inside and outside the bubble leads to interesting dynamics.
The fields reflected inside the bubble propagate, reaching the front boundary before the incident wave
packet has transited the bubble.  Consequently, near the front boundary of the bubble, reflected
fields due to the leading edge of the wave packet interfere
with the transmitted fields due to the trailing edge of the wave
packet. This interference pattern is discernible in Fig. 21xxx. 

After the original wave packet has propagated past the bubble, 
the fields are as in Fig. 22. There are two distinct
features which correspond to side and back scattering implying
that there are two internal reflections inside the bubble. At
this juncture, there are no electric field remnants left behind
in the bubble; thus, there will be no additional patterns for
side and back scattering. Along the same lines of discussion
for scattering by an elliptical dielectric, the scattering off the
bubble leads to the appearance of the y-component of the
magnetic field. The combination of Bx and By satisfies 
$\nabla \pmb{\cdot}  {\mathbf B} \left( x,y, t \right) =  0$.
Figures 23 and 24 show the two components of the
magnetic flux density after the wave packet has propagated
away from the bubble.

\begin{figure}[!t]
\centerline{\includegraphics[width=0.8 \columnwidth]{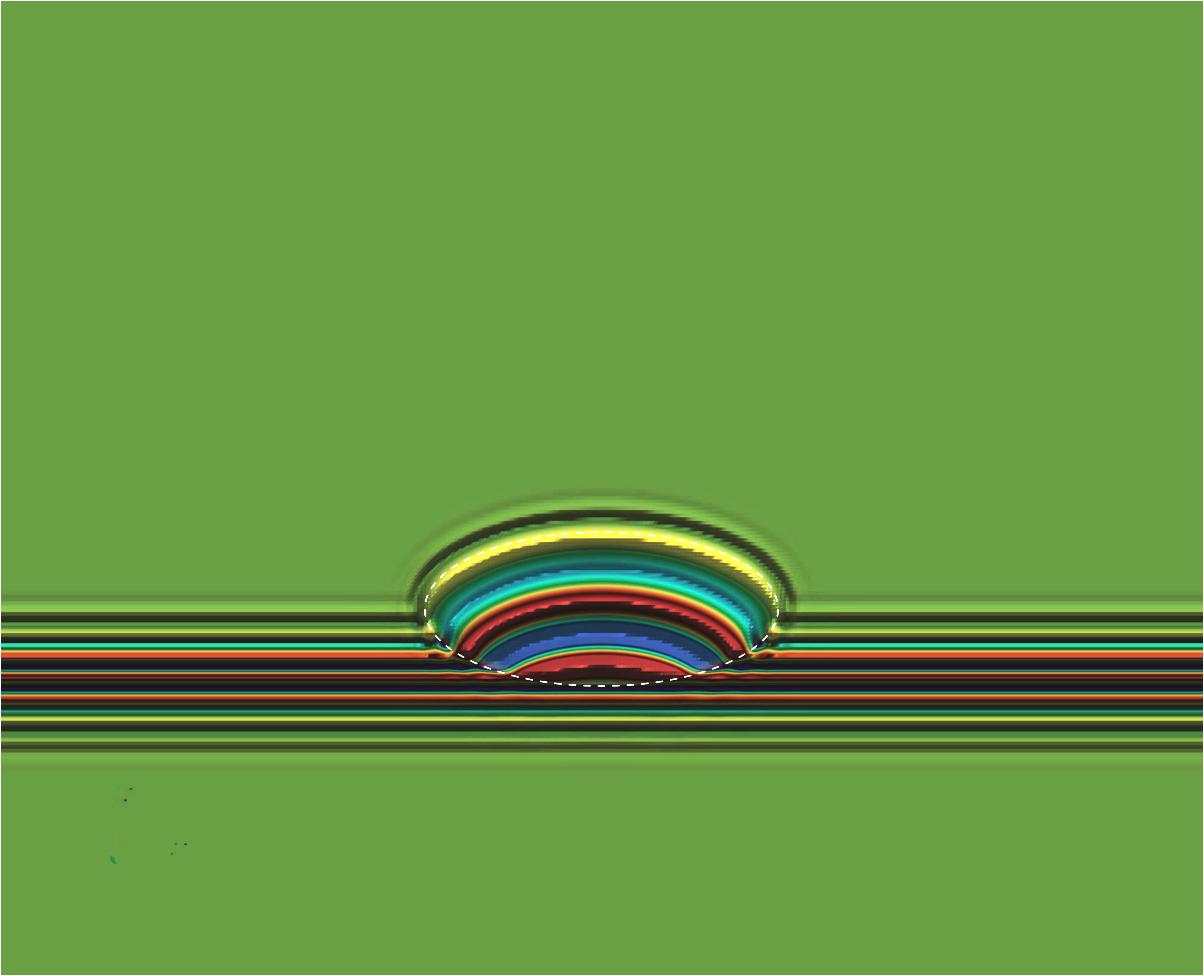}}
\caption{The propagation of wave fields speeds up inside the dielectric leading
to transmission of fields through the back end of the bubble ahead of the waves
in the dielectric. The reflected fields are propagating toward the front end. The
dashed surface marks the vacuum-dielectric interface.}
\label{fig19}
\end{figure}

\begin{figure}[!t]
\centerline{\includegraphics[width=0.8 \columnwidth]{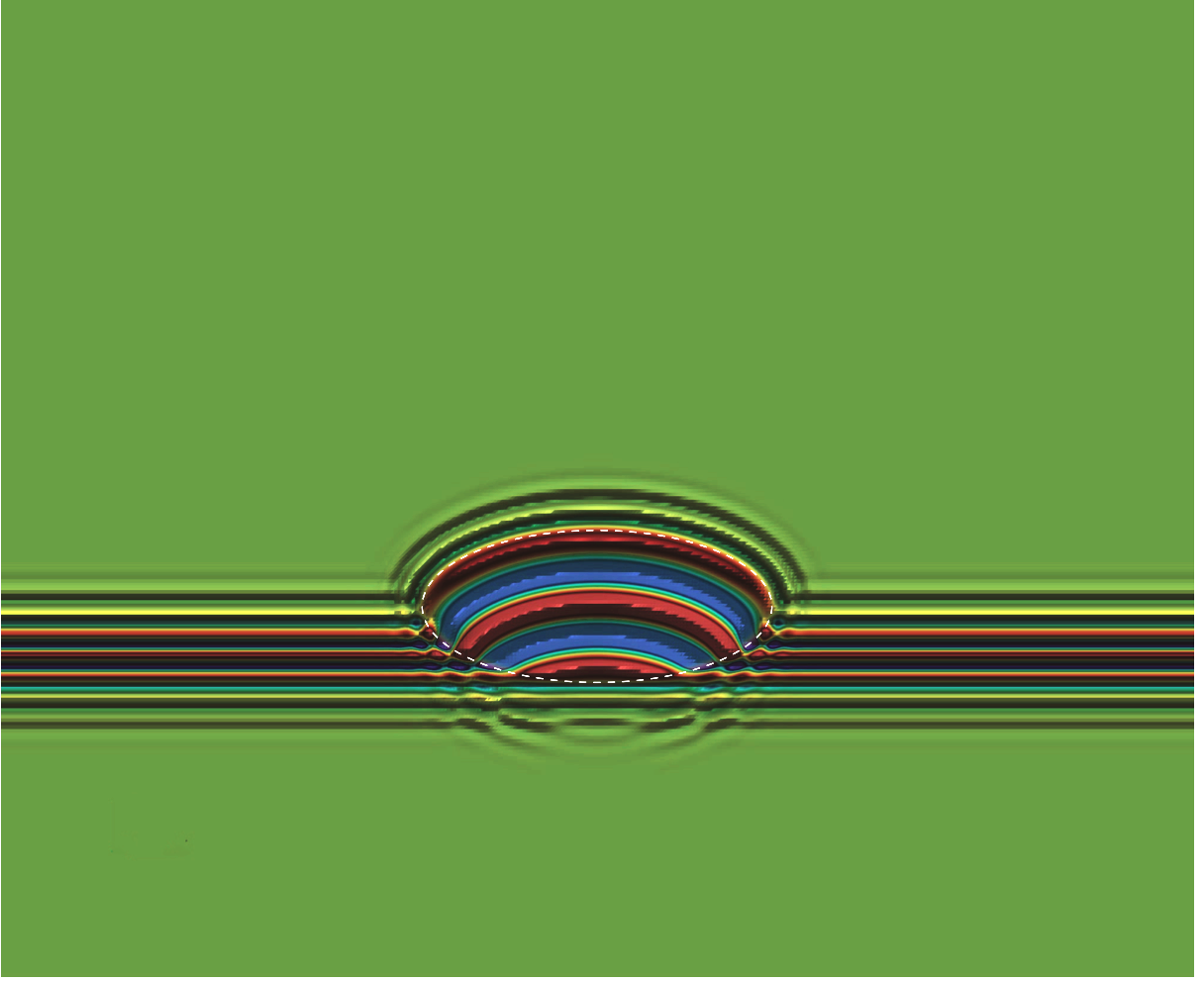}}
\caption{A short time later, internal fields reflected from the back end of
the bubble have transmitted through the front end leading to side and back
scattering. The field structures near the dielectric-vacuum interface are due
to topological mismatch between the planar wave fronts and the elliptical
boundary.}
\label{fig20}
\end{figure}

\begin{figure}[!t]
\centerline{\includegraphics[width=0.8 \columnwidth]{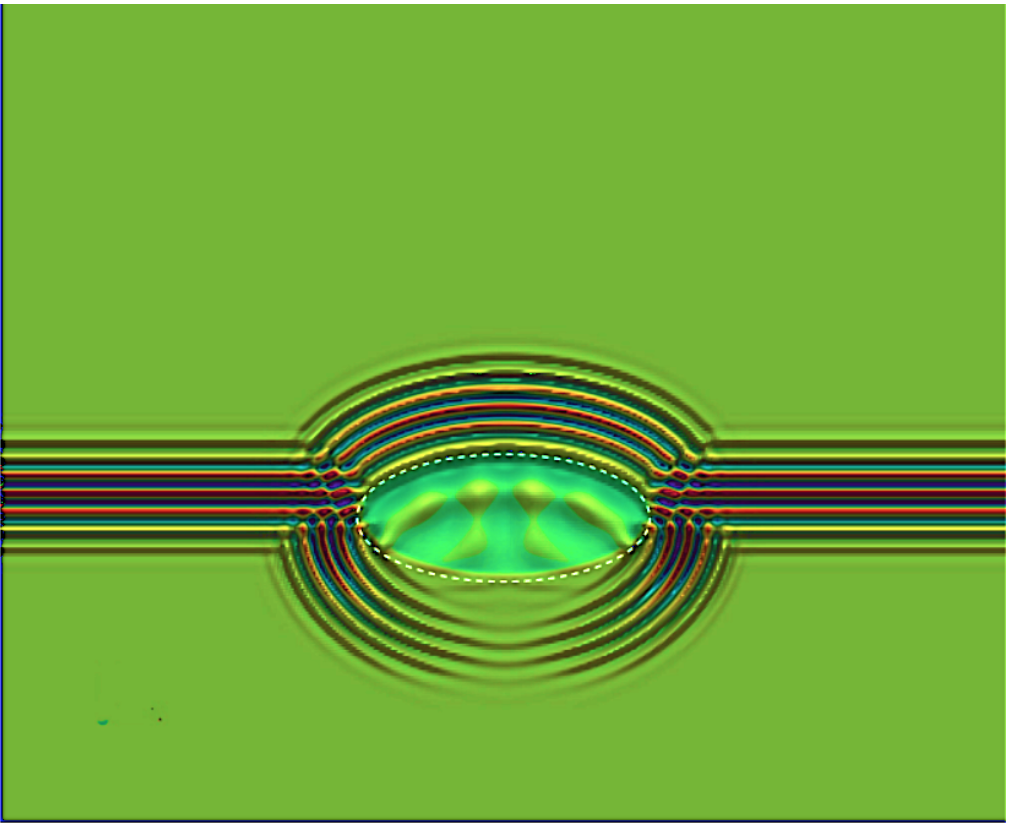}}
\caption{As the wave packet propagates past the bubble, the fields inside are
noticeable and are relatively weak. The wave pattern inside the bubble is due
to interference between forward and backward propagating waves. Since the
wave speed inside the bubble is three times that in the dielectric background,
the transmitted fields due to the leading part of the wave packet get partially
reflected from the back end of the bubble, and interfere with the transmitted
fields at the front end due to the trailing part of the wave packet.}
\label{fig21}
\end{figure}

\begin{figure}[!t]
\centerline{\includegraphics[width=0.8 \columnwidth]{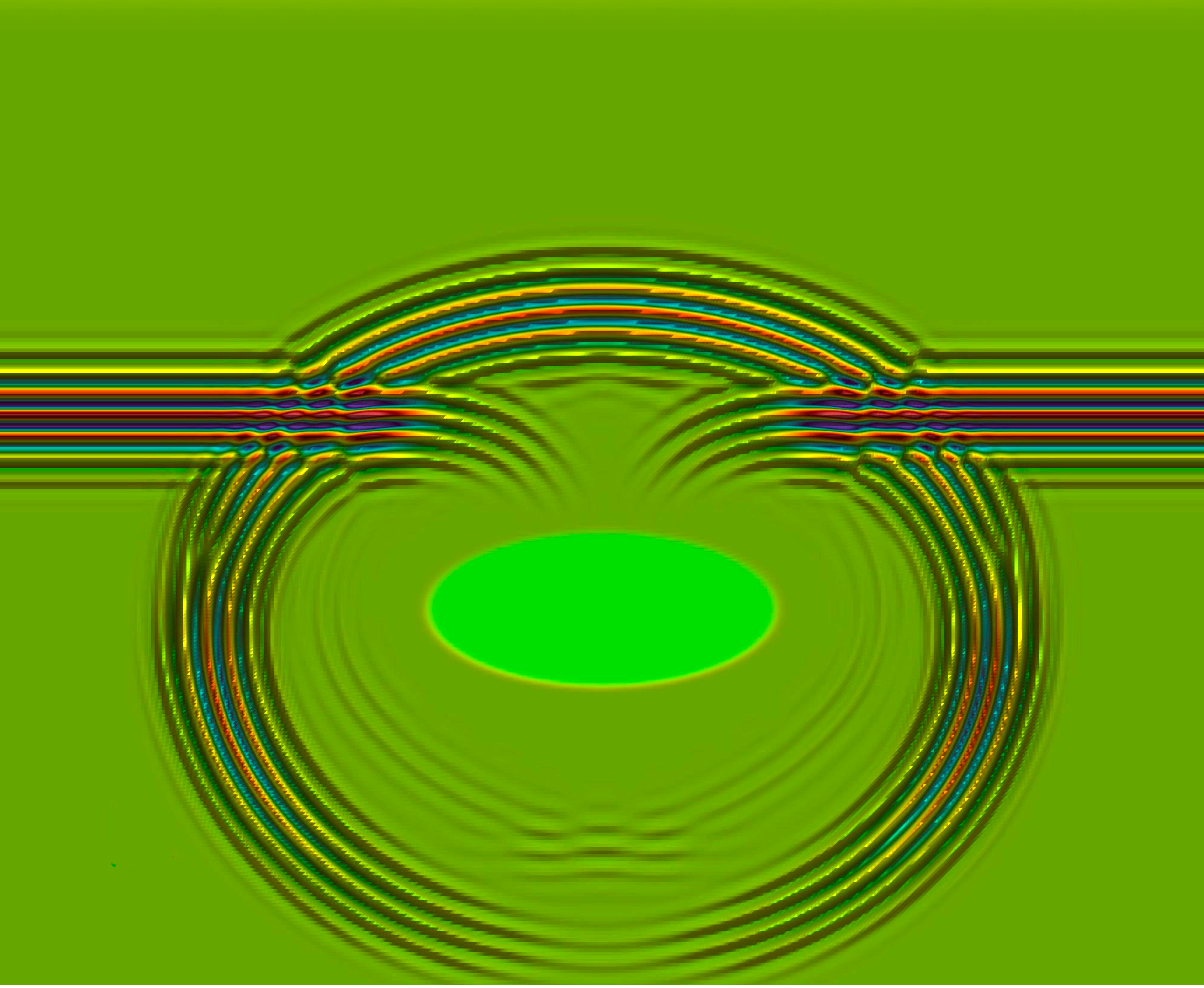}}
\caption{actually Fig 22.  The electric field profile after the wave packet has propagated past
the bubble. There are two stages leading to side and back scattering ? both,
due to internal reflections inside the bubble. At this juncture, wave fields are
propagating away from the bubble curtailing any additional scattering events.}
\label{fig22}
\end{figure}

\begin{figure}[!t]
\centerline{\includegraphics[width=0.8 \columnwidth]{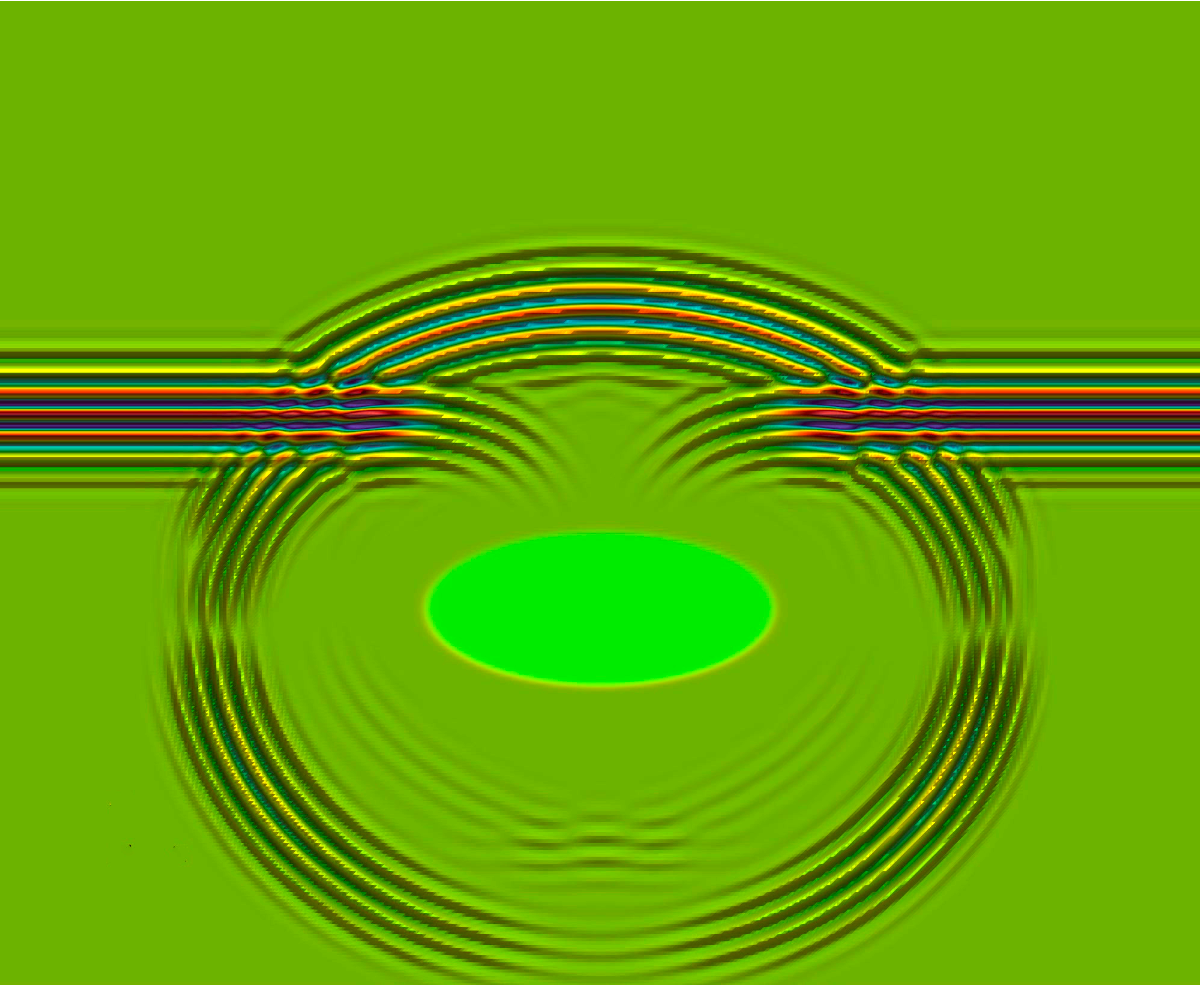}}
\caption{The corresponding $B_x$ field for the field $E_z$, Fig. 22}
\label{fig23}
\end{figure}

\begin{figure}[!t]
\centerline{\includegraphics[width=0.8 \columnwidth]{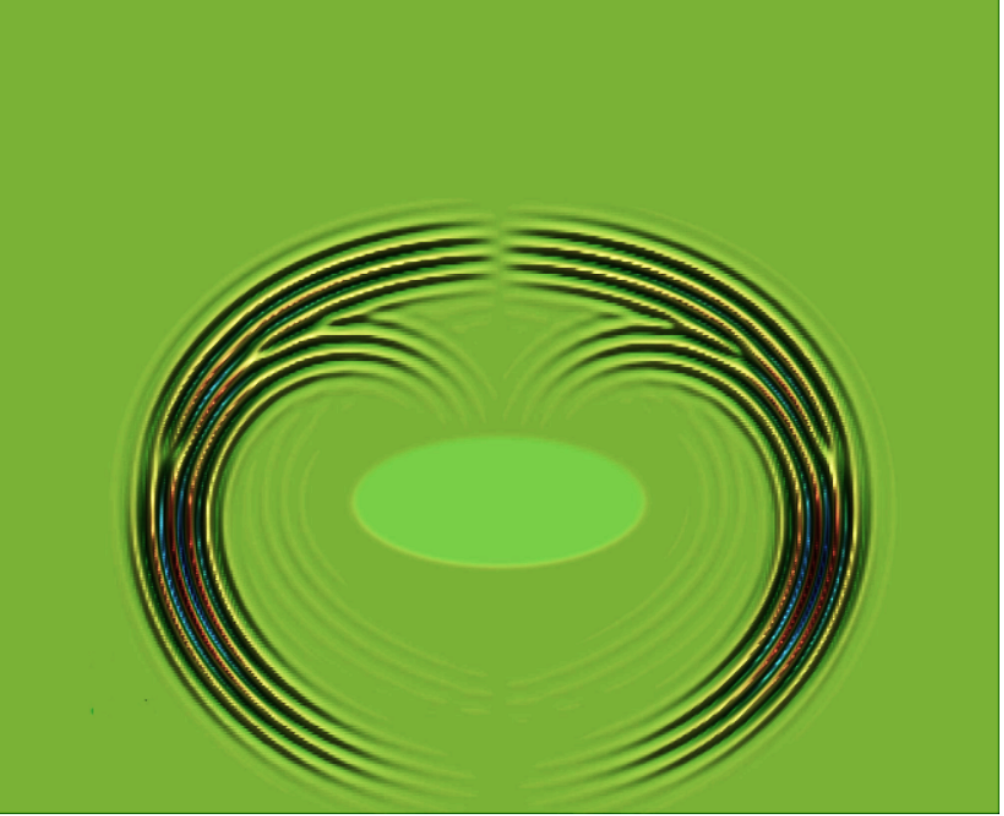}}
\caption{The corresponding self-consistently QLA generated $B_y$ field for the field $E_z$, Fig. 22.
The fields $B_x$ in Fig.23 and $B_y$ here in Fig. 24 will have $\nabla \cdot \bf{B} = 0$ to machine
accuracy.}
\label{fig24}
\end{figure}

\section{DISCUSSION}
A few aspects of transient scattering can be intuitively
understood using the Kirchhoff tangent plane approximation \cite{voro1999}.
In this approximation, the field at any point of the elliptic
scatterer is approximated to be the same as on the tangent
plane at that point. We are not going to concern ourselves with
the domain of validity of this approximation. Rather, we
want to gain simple insight into a complicated scattering
process resulting from simulations displayed in Section V. The
tangent plane separates two regions with different dielectric
permittivities; in particular, one region having the permittivity
of the background medium and the other region having the
permittivity of the scatterer. We will assume that the incoming
wave propagating in the background medium is a plane wave,
and its scattering at each point on the surface of the scatterer is
given by the conventional Fresnel reflection and transmission coefficients
for a smooth plane.  For an $s$-polarized plane wave incident at an angle
$\theta_i$ with respect to the normal to a plane surface, the respective
Fresnel reflectance and transmittance
\begin{equation}  \label{fresn} 
R = \left ( \frac{n_1 cos \theta_i - Re (\sqrt{n_2^2 - n_1^2 sin^2\theta_i})}{
 n_1 cos \theta_i +Re (\sqrt{n_2^2 - n_1^2 sin^2\theta_i})} \right )^2 , 
  T = 1 - R .
 \end{equation}
where $n_1$ and $n_2$ are the indices of refraction of the media
corresponding to the propagation of incident and transmitted
waves, respectively, and $Re$ denotes the real part. At the vertex
of an ellipse facing the incoming wave, the tangent plane is the $x-z$ plane
and the normal is in the $y$-direction.  Between the two co-vertices, the normal
spans a range $[-\pi/2,\pi/2]$; thus $\theta_i$ spans the same range. Since \eqref{fresn} is
symmetric about $\theta_i=0$, we need study the reflectance and transmittance for 
$0 \le \theta_i \le \pi/2$.  Figures 25  shows the Fresnel coefficients for $n_1=1$ and $n_2 = 3$ as
a function of $\theta_i$, while Fig 26 for the Fresnel coefficients for $n_1=3$ and $n_2 = 1$.

First, we apply the tangent plane approximation to the front
of the scatterer. For the dielectric ellipse, Fig. 25 shows that $R_1=T_1$
for  $\theta_i  \approx 0.336 \pi = \theta_{i1}$. For $\theta_i \ge \theta_{i1}$, $R_1 > T_1$; thus,
implying side-scattering which is observed in Figures 3, 4, and
5. For $\theta_i \le \theta_{i1}$, transmittance dominates and there is little to
no back scattering. In contrast, for the elliptic vacuum bubble, from
Fig. 26 we note that $R_2 = T_2$ when $\theta_i \approx 0.094 \pi = \theta_{i2}$ with
$T_2 > R_2$ for $\theta_1 < \theta_{i2}$. For angles $\theta_i > \theta_{i2}$, $R_2 > T_2$
with complete internal reflection for angles $\theta_i > \theta_{ic}  \approx 0.108 \pi$, 
the critical angle.  Again side scattering dominates
as seen in Figs. 20 and 21 while transmittance is prominent
for incidence angles close to normal.

Next, we apply the tangent plane approximation to the back
end of the scatterer. In this situation, Fig. 26 is pertinent to
propagation inside the dielectric scatterer while Fig. 25 applies to
scattering inside the bubble. Figures 7, 8, and 9 are a clear indication
 of total internal reflection occurring at points away from
the vertex in the back end of the dielectric scatterer; for angles
near the vertex the transmission of the fields dominates over
any reflections. For the bubble, Fig. 21 shows transmission
dominating over reflections in the forward direction and weak
internal reflections at angles away from the normal. This is in
broad agreement with the results in Fig. 26. Since there is lack
of total internal reflections, the wave fields inside the bubble are 
weak when compared to the fields inside the dielectric scatterer.

\begin{figure}[!t]
\centerline{\includegraphics[width= \columnwidth]{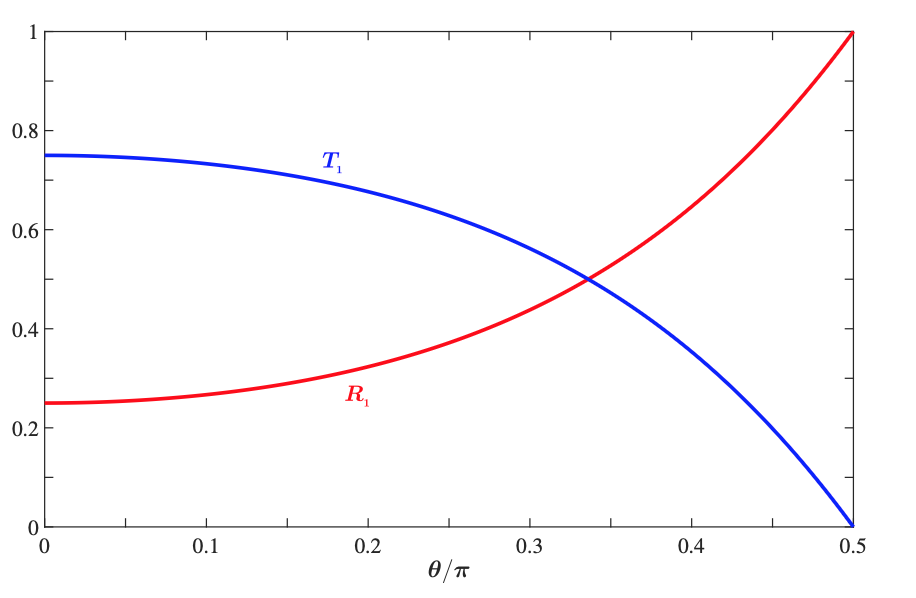}}
\caption{The reflectance $R_1$ and transmittance $T_1$ coefficients as a function
of the angle of incidence $\theta_i$ of a plane wave on a planar surface separating
two dielectric media. The results are for $n_1 = 1$ and $n_2 = 3.$}
\label{fig25}
\end{figure}

\begin{figure}[!t]
\centerline{\includegraphics[width= \columnwidth]{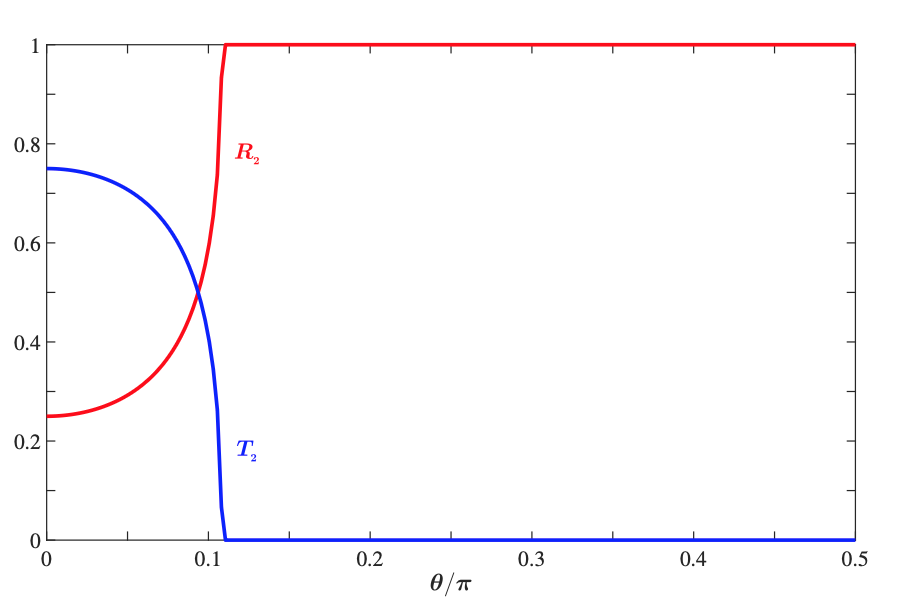}}
\caption{The reflectance $R_2$ and transmittance $T_2$ coefficients for $n_1 = 3$ and
$n_2 = 1$. For $\theta_i \ge 0.108 \pi$, the incident plane wave undergoes total internal
reflection.}
\label{fig26}
\end{figure}

\section{CONCLUSION}

A QLA for electromagnetic wave propagation in given linear dielectric media is presented that is based
on ideas from quantum computing :  can the time evolution of the  fields be determined from a sequence of
 unitary operators?  While theory indicates that there does exist such a fully unitary sequence of operators,
 the QLA model explored in this paper consists of 16 unitary collide-stream operators we do required the
 introduction of 2 non-unitary terms to recovery Maxwell equations to second order in the lattice mesh size.
 The beauty of a fully unitary QLA sequence is that it is immediately encodable on a quantum computer
 using 1-qubit and 2-qubit gates, and the total electromagnetic energy is just the norm of the Dyson variables 
 and so is exactly conserved to machine accuracy throughout the entire simulation run.  Our current QLA with
 16 unitary out of 18  operators for a time advancement yields energy conservation to seven significant figures. 
 It should be noted that the time evolution plots of the electromagnetic field from our initial value code would 
 not be easily available from an actual quantum computer run.  Rather, one would output reduced data like the
 energy distribution and the Fresnel coefficients but which would still provide insights on the underlying
 physical mechanisms that are occurring.  Even as a classical algorithm, QLA  has important implications for
 being run on classical supercomputers for its extreme parallelization to the total number of cores available.
 This conclusion also holds for QLA simulations for nonlinear problems (e.g., spinor BEC turbulence) since
 the nonlinear terms are only algebraic in QLA rather than involving nonlinear convective derivatives as in
 standard CFD. 
 
QLA yields the transient fields captured in our initial value code and so helps unravel some of the
intricacies of a spatially confined wave packet.  We find the back
scattering of the wave packet is a secondary effect driven by
the fields which are temporarily confined within the dielectric scatterer.
The side scattering and forward scattering tends to dominate over backscattering.
 Some of the basic physics
leading to the scattering results can be tacitly understood using
the Kirchhoff tangent plane approximation. It helps explain the
difference between scattering by a dielectric ellipse surrounded
by vacuum and a vacuum bubble embedded in a dielectric.

Future work will be to uncover a fully unitary QLA for Maxwell equations in linear dielectric media.
 
\section*{ACKNOWLEDGMENTS}
This research was partially supported by Department of Energy grants DE-SC0021647, DE-FG02-
91ER-54109, DE-SC0021651, DE-SC0021857, and DE-SC0021653. This work has been carried out
within the framework of the EUROfusion Consortium, funded by the European Union via the
Euratom Research and Training Programme (Grant Agreement No. 101052200 - EUROfusion).
Views and opinions expressed, however, are those of the authors only and do not necessarily reflect
those of the European Union or the European Commission. Neither the European Union nor the
European Commission can be held responsible for them. E. K. is supported by the Basic Research
Program, NTUA, PEVE. K.H is supported by the National Program for Controlled Thermonuclear
Fusion, Hellenic Republic. 
This research used resources of the National Energy Research Scientific Computing
 Center (NERSC), a U.S. Department of Energy  Science User Facility located at
Lawrence Berkeley National Laboratory, operated under Contract No. DE-AC02-05CH11231 using
NERSC award FES-ERCAP0031599.

%\section*{REFERENCES}

\end{document}